\documentclass[final,twoside,11pt]{entics}
\usepackage{enticsmacro}
\usepackage{graphicx}
\usepackage[all]{xy}
\usepackage{hyperref}
\usepackage{amsmath,amssymb}
\usepackage{iftex}
\ifPDFTeX
  \usepackage[T1]{fontenc}
  \usepackage[utf8]{inputenc}
  \usepackage{textcomp} % provide euro and other symbols
\else % if luatex or xetex
  \usepackage{unicode-math} % this also loads fontspec
  \defaultfontfeatures{Scale=MatchLowercase}
  \defaultfontfeatures[\rmfamily]{Ligatures=TeX,Scale=1}
\fi
\usepackage{lmodern}
\ifPDFTeX\else
\fi
\IfFileExists{upquote.sty}{\usepackage{upquote}}{}
\IfFileExists{microtype.sty}{% use microtype if available
  \usepackage[]{microtype}
  \UseMicrotypeSet[protrusion]{basicmath} % disable protrusion for tt fonts
}{}
\makeatletter
\@ifundefined{KOMAClassName}{% if non-KOMA class
  \IfFileExists{parskip.sty}{%
    \usepackage{parskip}
  }{% else
    \setlength{\parindent}{0pt}
    \setlength{\parskip}{6pt plus 2pt minus 1pt}}
}{% if KOMA class
  \KOMAoptions{parskip=half}}
\makeatother
\usepackage{xcolor}
\usepackage{color}
\usepackage{fancyvrb}

\DefineVerbatimEnvironment{Highlighting}{Verbatim}{commandchars=\\\{\}}
\definecolor{shadecolor}{RGB}{255,255,255}
\newenvironment{Shaded}{}{}

\newcommand{\DataTypeTok}[1]{\textcolor[rgb]{0.00,0.34,0.68}{#1}}

\newcommand{\KeywordTok}[1]{\textcolor[rgb]{0.12,0.11,0.11}{\textbf{#1}}}
\newcommand{\NormalTok}[1]{\textcolor[rgb]{0.12,0.11,0.11}{#1}}

\newcommand{\OtherTok}[1]{\textcolor[rgb]{0.00,0.43,0.16}{#1}}
\newcommand{\PreprocessorTok}[1]{\textcolor[rgb]{0.00,0.43,0.16}{#1}}

\providecommand{\tightlist}{%
  \setlength{\itemsep}{0pt}\setlength{\parskip}{0pt}}
\usepackage{quiver}
\usepackage[utf8]{inputenc}
\usepackage{newunicodechar}
\newunicodechar{λ}{\ensuremath{\mathnormal\lambda}}
\newunicodechar{∀}{\ensuremath{\mathnormal\forall}}
\newunicodechar{≡}{\ensuremath{\mathnormal\equiv}}
\newunicodechar{ℓ}{\ensuremath{\mathnormal\ell}}
\newunicodechar{κ}{\ensuremath{\mathnormal\kappa}}
\newunicodechar{Σ}{\ensuremath{\mathnormal\Sigma}}
\newunicodechar{⊔}{\ensuremath{\mathnormal\sqcup}}
\newunicodechar{♭}{\ensuremath{\mathnormal\flat}}
\newunicodechar{ε}{\ensuremath{\mathnormal\epsilon}}
\newunicodechar{₀}{\ensuremath{\mathnormal{_0}}}
\newunicodechar{⊥}{\ensuremath{\mathnormal\bot}}
\newunicodechar{⊤}{\ensuremath{\mathnormal\top}}
\newunicodechar{α}{\ensuremath{\mathnormal\alpha}}
\newunicodechar{β}{\ensuremath{\mathnormal\beta}}
\newunicodechar{η}{\ensuremath{\mathnormal\eta}}
\newunicodechar{⁻}{\ensuremath{\mathnormal{^-}}}
\newunicodechar{¹}{\ensuremath{\mathnormal{^1}}}
\newunicodechar{ℕ}{\ensuremath{\mathbb{N}}}
\newunicodechar{ω}{\ensuremath{\mathnormal\omega}}
\DeclareMathAlphabet{\mathcal}{OMS}{cmsy}{m}{n}
\ifLuaTeX
  \usepackage{selnolig}  % disable illegal ligatures
\fi
\usepackage{bookmark}
\IfFileExists{xurl.sty}{\usepackage{xurl}}{} % add URL line breaks if available
\hypersetup{
  pdftitle={Parametricity via Cohesion},
  pdfauthor={C.B. Aberlé},% hidelinks,
  pdfcreator={LaTeX via pandoc}}
\def\conf{MFPS 2024}
\def\myconf{mfps40} 	%%Fill in the acronym for your conference (with year)
\volume{4}			%Fill in the ENTICS volume number here
\def\volu{4}			% and here
\def\papno{1}			%Fill in your paper number here
\def\mydoi{14710}%
\def\lastname{Aberle} %Lastnames appear in the running header 
\def\runtitle{Parametricity via Cohesion} %Short title appears in the running header on even pages. 
\def\copynames{C. Aberlé}    %% Fill in the first initial and last name of the authors
\begin{document}
%%%Note the beginning and end of the frontmatter section that starts here%%%%%
\begin{frontmatter}
  \title{Parametricity via Cohesion} 						%%Title here and the
  %%Text of \thanks[ALL} here..
 %%%%%%%%%%%%%%%%%%%%%%%%%%%%			This Thanks is optional.
  %%%%Now the author(s) names(s)%%%%%
  \author{C.B. Aberlé}	%%Note NO SPACE between 
   		%last name and \thanksref{...} 
    %%%Next come the addresses%%%%
   \address{Computer Science Department\\ Carnegie Mellon University\\				%or between \thanksrefs...
    Pittsburgh, PA, USA}  							
   \thanks{Email: \href{mailto:caberle@andrew.cmu.edu} {\texttt{\normalshape
        caberle@andrew.cmu.edu}}} 
   %%%Note: if both authors share same institution, only list the address once, after the second 
   %%%author. 
   %%%There also is a link from the first author to the co-author's address to show how to list 
   %%%affiliations to more than one institution, when needed. 
\begin{abstract}
Parametricity is a key metatheoretic property of type
systems, which implies strong uniformity \& modularity
properties of the structure of types within systems possessing it. In recent
years, various systems of dependent type theory have emerged with the
aim of expressing such parametric reasoning in their internal
logic, toward the end of solving various problems arising from the complexity of higher-dimensional \emph{coherence conditions} in type theory. This paper presents a first step toward the
unification, simplification, and extension of these various methods for
internalizing parametricity. Specifically, I argue that there is an
essentially modal aspect of parametricity, which is intimately connected
with the category-theoretic concept of \emph{cohesion}. On this basis, I describe a general categorical semantics for modal
parametricity, develop a corresponding framework of axioms (with
computational interpretations) in dependent type theory that can be used
to internally represent and reason about such parametricity, and show
this in practice by implementing these axioms in Agda and using them to
verify parametricity theorems therein. I then demonstrate the utility of these axioms in managing the complexity of higher-dimensional coherence by deriving induction principles for higher inductive types, and in closing, I sketch the outlines
of a more general synthetic theory of parametricity, with applications in domains ranging from homotopy type theory to
the analysis of program modules.
\end{abstract}
\begin{keyword}
  parametricity, cohesion, type theory, category theory, homotopy type theory, constructive mathematics, agda.
\end{keyword}
\end{frontmatter}

\section{The past, present \& future of parametricity in type
theory}\label{the-past-present-future-of-parametricity-in-type-theory}

Reynolds \cite{Reynolds1983} began his seminal introduction of the concept of
\emph{parametricity} with the declaration that ``type structure is a
syntactic discipline for enforcing levels of abstraction.'' In the
ensuing decades, this idea of Reynolds' has been overwhelmingly
vindicated by the success of type systems in achieving modularity and
abstraction in domains ranging from programming to interactive theorem
proving. Yet the fate of Reynolds' particular strategy for formalizing
this idea is rather more ambiguous.

Reynolds' original analysis of parametricity targeted the polymorphic
\(\lambda\)-calculus (aka System F). Intuitively, polymorphic programs
in System F cannot inspect the types over which they are defined and so
must behave \emph{essentially the same} for all types at which they are
instantiated. To make this intuitive idea precise, Reynolds posed an
ingenious solution in terms of logical relations, whereby every System F
type is equipped with a suitable binary relation on its inhabitants,
such that all terms constructible in System F must preserve the
relations defined on their component types. On this basis, Reynolds was
able to establish many significant properties of the abstraction
afforded by System F's type structure, e.g: all closed terms of type
\(\forall X . X \to X\) are extensionally equivalent to the identity
function.

Reynolds' results have subsequently been extended to many type systems.
However, as the complexity and expressivity of these systems increases,
so too does the difficulty of defining parametricity relations for their
types, and proving that the inhabitants of these types respect their
corresponding relations. Moreover, the abstract, mathematical patterns
underlying the definitions of such relations have not always been clear.
The problem essentially is that the usual theory of parametricity
relations for type systems is ``analytic,'' i.e.~defined in terms of the
particular constructs afforded by those systems, when what would be more
desirable is a ``synthetic'' theory that boils the requirements for
parametricity down to a few axioms.

The need for such an axiomatic specification of parametricity is made
all the more apparent by recent developments in \emph{Homotopy Type
Theory} (HoTT) and related fields, wherein dependent type theories capable of
proving parametricity theorems \emph{internally} for their own type
structures have been developed and applied by various authors to solve
some major open problems in these fields (c.f. \cite{Cavallo2020}, \cite{kolomatskaia2024}). Each such type theory has
posed its own solution to the problem of internalizing parametricity,
and although some commonalities exist between these systems, there is as
yet no one framework that subsumes them all. In particular, the majority
of these type theories have targeted only specific semantic models
(usually some appropriately-chosen presheaf categories) rather than an
axiomatically-defined \emph{class} of models, and in this sense the
analysis of parametricity offered by these type theories remains
\emph{analytic,} rather than \emph{synthetic}. This in particular limits
any insight into whether and how these approaches to internal
parametricity may be related to one another, generalized, or simplified.

As a first step toward the unification, simplification, and
generalization of these approaches to internal parametricity, I propose
an analysis of parametricity in terms of the category-theoretic concept
of \emph{cohesion}. Cohesion here refers to a particular relation
between categories whereby one is equipped with an adjoint string of
modalities that together make its objects behave like abstract spaces,
whose points are bound together by some sort of \emph{cohesion} that
serves to constrain the maps that exist between these spaces. Such a
notion of cohesion is reminiscent of the informal idea behind Reynolds'
formulation of parametricity outlined above, particularly if one
interprets the \emph{cohesion} that binds types together as the
structure of relations that inhere between them. Moreover, by inspection
of the categorical models of extant type theories for internal
parametricity, along with classical models of parametricity, one finds
that many (if not quite all) of these exhibit cohesion, in this sense.
The main contribution of this paper is thus twofold: 1) to show that this basic
setup of cohesion is essentially \emph{all} that is needed to recover
classical parametricity results internally in dependent type theory, and 2) to show how this axiomatic framework can be applied in solving problems arising from the complexity of \emph{coherence conditions} in HoTT -- specifically the problem of deriving induction principles for higher inductive types from their recursors. As a further illustration and verification of this idea, this paper is also a
literate Agda document wherein the axioms for and theorems resulting
from such parametricity via cohesion have been fully formalized and
checked for validity.\footnote{The full source of this file can be found at:\\ {\href{https://github.com/cbaberle/Parametricity-via-Cohesion/blob/main/parametricity-via-cohesion.lagda.md}{https://github.com/cbaberle/Parametricity-via-Cohesion/blob/main/parametricity-via-cohesion.lagda.md}}}

\begin{Shaded}
\begin{Highlighting}[]
\PreprocessorTok{\{{-}\# OPTIONS {-}{-}rewriting {-}{-}cohesion {-}{-}flat{-}split {-}{-}without{-}K \#{-}\}}
\KeywordTok{module}\NormalTok{ parametricity{-}via{-}cohesion }\KeywordTok{where}
\end{Highlighting}
\end{Shaded}

\section{Cohesion \& Parametricity}\label{cohesion-and-parametricity}

The notion of \emph{cohesion} as an abstract characterization of when
one category (specifically a topos) behaves like a category of spaces
defined over the objects of another, is due primarily to Lawvere \cite{LawvereCohesion,LawvereGraph}. The
central concept of axiomatic cohesion is an arrangement of four adjoint
functors as in the following diagram:\[
\begin{tikzcd}
    {\mathcal{E}} \\
    \\
    {\mathcal{S}}
    \arrow[""{name=0, anchor=center, inner sep=0}, "\Gamma"{description}, curve={height=-12pt}, from=1-1, to=3-1]
    \arrow[""{name=1, anchor=center, inner sep=0}, "\nabla"{description}, curve={height=30pt}, from=3-1, to=1-1]
    \arrow[""{name=2, anchor=center, inner sep=0}, "\Delta"{description}, curve={height=-12pt}, from=3-1, to=1-1]
    \arrow[""{name=3, anchor=center, inner sep=0}, "\Pi"{description}, curve={height=30pt}, from=1-1, to=3-1]
    \arrow["\dashv"{anchor=center}, draw=none, from=3, to=2]
    \arrow["\dashv"{anchor=center}, draw=none, from=2, to=0]
    \arrow["\dashv"{anchor=center}, draw=none, from=0, to=1]
\end{tikzcd}
\] where \(\mathcal{E,S}\) are both topoi, \(\Delta, \nabla\) are both
fully faithful, and \(\Pi\) preserves finite products. Given such an
arrangement, we think of the objects of \(\mathcal{E}\) as \emph{spaces}
and those of \(\mathcal{S}\) as \emph{sets} (even if \(\mathcal{S}\) is
not the category of sets), where \(\Gamma\) is the functor that sends a
space to its set of points, \(\Delta\) sends a set to the corresponding
\emph{discrete space}, \(\nabla\) sends a set to the corresponding
\emph{codiscrete space}, and \(\Pi\) sends a space to its set of
connected components. These in turn induce a string of adjoint
modalities on \(\mathcal{E}\): \[
\smallint \dashv \flat \dashv \sharp 
\] where \(\smallint = \Delta \circ \Pi\) and
\(\sharp = \nabla \circ \Gamma\) are idempotent monads, and
\(\flat = \Delta \circ \Gamma\) is an idempotent comonad.

A concrete example of cohesion comes from the category of reflexive
graphs \(\mathbf{RGph}\), which is cohesive over the category of sets
\(\mathbf{Set}\) \cite{LawvereGraph}. Here, \(\Gamma\) is the functor that sends a reflexive
graph to its set of vertices, \(\Delta\) sends a set \(V\) to the
``discrete'' reflexive graph on \(V\) whose only edges are self-loops,
\(\nabla\) sends \(V\) to the ``codiscrete'' graph where there is a
unique edge between any pair of vertices, and \(\Pi\) sends a reflexive
graph to its set of (weakly) connected components. It is worth noting,
at this point, that many classical models of parametricity (e.g. \cite{atkey}) are based
upon semantic interpretations of type structure in terms of reflexive graphs. This, I wish
to argue, is no accident, and the key property of reflexive graphs
underlying such interpretations is precisely their cohesive structure.
More generally, for any base topos \(\mathcal{S}\), we may construct its
corresponding topos \(\mathbf{RGph}(\mathcal{S})\) of \emph{internal
reflexive graphs}, which will similarly be cohesive over
\(\mathcal{S}\), so we can in fact use the language of such internal
reflexive graphs to derive parametricity results for \emph{any} topos
(or indeed, any \(\infty\)-topos, as described below).

In fact, this same setup of cohesion is interpretable, \emph{mutatis
mutandis}, in the case where \(\mathcal{E,S}\) are not topoi, but rather
\emph{\(\infty\)-topoi}, i.e.~models of homotopy type theory \cite{Shulman2018}. This
allows us to use the language of homotopy type theory -- suitably
extended with constructs for the above-described modalities (the
\(\flat\) modality in particular, which, for technical reasons, cannot
be axiomatized directly in ordinary HoTT) -- to work
\emph{synthetically} with the structure of such a cohesive
\(\infty\)-topos. For present purposes, we accomplish this by working in
Agda with the \texttt{-\/-cohesion} and \texttt{-\/-flat-split} flags
enabled, along with \texttt{-\/-without-K}, which ensures compatibility
with the treatment of propositional equality in HoTT.

I therefore begin by recalling some standard definitions from HoTT \cite{hottbook},
which shall be essential in defining much of the structure to follow.
Essentially all of these definitions have to do with the \emph{identity
type} former \texttt{\_≡\_} and its associated constructor
\texttt{refl\ :\ ∀\ \{ℓ\}\ \{A\ :\ Set\ ℓ\}\ \{a\ :\ A\}\ →\ a\ ≡\ a},
as defined in the module \texttt{Agda.Builtin.Equality}:

\begin{Shaded}
\begin{Highlighting}[]
\KeywordTok{module}\NormalTok{ hott }\KeywordTok{where}
\end{Highlighting}
\end{Shaded}

First of all, we have the induction principle for the identity type, aka
the \texttt{J} rule:

\begin{Shaded}
\begin{Highlighting}[]
\NormalTok{    J }\OtherTok{:} \OtherTok{∀} \OtherTok{\{}\NormalTok{ℓ κ}\OtherTok{\}} \OtherTok{\{}\NormalTok{A }\OtherTok{:} \DataTypeTok{Set}\NormalTok{ ℓ}\OtherTok{\}} \OtherTok{\{}\NormalTok{a }\OtherTok{:}\NormalTok{ A}\OtherTok{\}}
        \OtherTok{→} \OtherTok{(}\NormalTok{B }\OtherTok{:} \OtherTok{(}\NormalTok{b }\OtherTok{:}\NormalTok{ A}\OtherTok{)} \OtherTok{→}\NormalTok{ a ≡ b }\OtherTok{→} \DataTypeTok{Set}\NormalTok{ κ}\OtherTok{)}
        \OtherTok{→} \OtherTok{\{}\NormalTok{b }\OtherTok{:}\NormalTok{ A}\OtherTok{\}} \OtherTok{→} \OtherTok{(}\NormalTok{p }\OtherTok{:}\NormalTok{ a ≡ b}\OtherTok{)} \OtherTok{→}\NormalTok{ B a refl }\OtherTok{→}\NormalTok{ B b p}
\NormalTok{    J B refl b }\OtherTok{=}\NormalTok{ b}
\end{Highlighting}
\end{Shaded}

We then obtain the operation of \emph{transport} as the recursor for the
identity type:

\begin{Shaded}
\begin{Highlighting}[]
\NormalTok{    transp }\OtherTok{:} \OtherTok{∀} \OtherTok{\{}\NormalTok{ℓ κ}\OtherTok{\}} \OtherTok{\{}\NormalTok{A }\OtherTok{:} \DataTypeTok{Set}\NormalTok{ ℓ}\OtherTok{\}} \OtherTok{\{}\NormalTok{a b }\OtherTok{:}\NormalTok{ A}\OtherTok{\}} 
             \OtherTok{→} \OtherTok{(}\NormalTok{B }\OtherTok{:}\NormalTok{ A }\OtherTok{→} \DataTypeTok{Set}\NormalTok{ κ}\OtherTok{)} \OtherTok{→} \OtherTok{(}\NormalTok{a ≡ b}\OtherTok{)} \OtherTok{→}\NormalTok{ B a }\OtherTok{→}\NormalTok{ B b}
\NormalTok{    transp B p b }\OtherTok{=}\NormalTok{ J }\OtherTok{(λ}\NormalTok{ a }\OtherTok{\_} \OtherTok{→}\NormalTok{ B a}\OtherTok{)}\NormalTok{ p b}
\end{Highlighting}
\end{Shaded}

Additionally, both \texttt{J} and \texttt{transp} are symmetric, and so
can be applied ``in the opposite direction'':

\begin{Shaded}
\begin{Highlighting}[]
\NormalTok{    J⁻¹ }\OtherTok{:} \OtherTok{∀} \OtherTok{\{}\NormalTok{ℓ κ}\OtherTok{\}} \OtherTok{\{}\NormalTok{A }\OtherTok{:} \DataTypeTok{Set}\NormalTok{ ℓ}\OtherTok{\}} \OtherTok{\{}\NormalTok{a }\OtherTok{:}\NormalTok{ A}\OtherTok{\}}
          \OtherTok{→} \OtherTok{(}\NormalTok{B }\OtherTok{:} \OtherTok{(}\NormalTok{b }\OtherTok{:}\NormalTok{ A}\OtherTok{)} \OtherTok{→}\NormalTok{ a ≡ b }\OtherTok{→} \DataTypeTok{Set}\NormalTok{ κ}\OtherTok{)}
          \OtherTok{→} \OtherTok{\{}\NormalTok{b }\OtherTok{:}\NormalTok{ A}\OtherTok{\}} \OtherTok{→} \OtherTok{(}\NormalTok{p }\OtherTok{:}\NormalTok{ a ≡ b}\OtherTok{)} \OtherTok{→}\NormalTok{ B b p }\OtherTok{→}\NormalTok{ B a refl}
\NormalTok{    J⁻¹ B refl b }\OtherTok{=}\NormalTok{ b}

\NormalTok{    transp⁻¹ }\OtherTok{:} \OtherTok{∀} \OtherTok{\{}\NormalTok{ℓ κ}\OtherTok{\}} \OtherTok{\{}\NormalTok{A }\OtherTok{:} \DataTypeTok{Set}\NormalTok{ ℓ}\OtherTok{\}} \OtherTok{\{}\NormalTok{a b }\OtherTok{:}\NormalTok{ A}\OtherTok{\}} 
               \OtherTok{→} \OtherTok{(}\NormalTok{B }\OtherTok{:}\NormalTok{ A }\OtherTok{→} \DataTypeTok{Set}\NormalTok{ κ}\OtherTok{)} \OtherTok{→} \OtherTok{(}\NormalTok{a ≡ b}\OtherTok{)} \OtherTok{→}\NormalTok{ B b }\OtherTok{→}\NormalTok{ B a}
\NormalTok{    transp⁻¹ B p b }\OtherTok{=}\NormalTok{ J⁻¹ }\OtherTok{(λ}\NormalTok{ a }\OtherTok{\_} \OtherTok{→}\NormalTok{ B a}\OtherTok{)}\NormalTok{ p b}
\end{Highlighting}
\end{Shaded}

Moreover, since all functions must preserve relations of identity, we
may apply a function to both sides of an identification as follows:

\begin{Shaded}
\begin{Highlighting}[]
\NormalTok{    ap }\OtherTok{:} \OtherTok{∀} \OtherTok{\{}\NormalTok{ℓ κ}\OtherTok{\}} \OtherTok{\{}\NormalTok{A }\OtherTok{:} \DataTypeTok{Set}\NormalTok{ ℓ}\OtherTok{\}} \OtherTok{\{}\NormalTok{B }\OtherTok{:} \DataTypeTok{Set}\NormalTok{ κ}\OtherTok{\}} \OtherTok{\{}\NormalTok{a b }\OtherTok{:}\NormalTok{ A}\OtherTok{\}}
         \OtherTok{→} \OtherTok{(}\NormalTok{f }\OtherTok{:}\NormalTok{ A }\OtherTok{→}\NormalTok{ B}\OtherTok{)} \OtherTok{→}\NormalTok{ a ≡ b }\OtherTok{→}\NormalTok{ f a ≡ f b}
\NormalTok{    ap f refl }\OtherTok{=}\NormalTok{ refl}
\end{Highlighting}
\end{Shaded}

The notion of \emph{contractibility} then expresses the idea that a type
is essentially uniquely inhabited.

\begin{Shaded}
\begin{Highlighting}[]
\NormalTok{    isContr }\OtherTok{:} \OtherTok{∀} \OtherTok{\{}\NormalTok{ℓ}\OtherTok{\}} \OtherTok{(}\NormalTok{A }\OtherTok{:} \DataTypeTok{Set}\NormalTok{ ℓ}\OtherTok{)} \OtherTok{→} \DataTypeTok{Set}\NormalTok{ ℓ}
\NormalTok{    isContr A }\OtherTok{=}\NormalTok{ Σ A }\OtherTok{(λ}\NormalTok{ a }\OtherTok{→} \OtherTok{(}\NormalTok{b }\OtherTok{:}\NormalTok{ A}\OtherTok{)} \OtherTok{→}\NormalTok{ a ≡ b}\OtherTok{)}
\end{Highlighting}
\end{Shaded}

Similarly, the notion of \emph{equivalence} expresses the idea that a
function between types has an \emph{essentially unique}
inverse.\footnote{Those familiar with HoTT may note that I use the
  \emph{contractible fibres} definition of equivalence, as this shall be
  the most convenient to work with, for present purposes.}

\begin{Shaded}
\begin{Highlighting}[]
\NormalTok{    isEquiv }\OtherTok{:} \OtherTok{∀} \OtherTok{\{}\NormalTok{ℓ κ}\OtherTok{\}} \OtherTok{\{}\NormalTok{A }\OtherTok{:} \DataTypeTok{Set}\NormalTok{ ℓ}\OtherTok{\}} \OtherTok{\{}\NormalTok{B }\OtherTok{:} \DataTypeTok{Set}\NormalTok{ κ}\OtherTok{\}} 
              \OtherTok{→} \OtherTok{(}\NormalTok{A }\OtherTok{→}\NormalTok{ B}\OtherTok{)} \OtherTok{→} \DataTypeTok{Set} \OtherTok{(}\NormalTok{ℓ ⊔ κ}\OtherTok{)}
\NormalTok{    isEquiv }\OtherTok{\{}\NormalTok{A }\OtherTok{=}\NormalTok{ A}\OtherTok{\}} \OtherTok{\{}\NormalTok{B }\OtherTok{=}\NormalTok{ B}\OtherTok{\}}\NormalTok{ f }\OtherTok{=} 
        \OtherTok{(}\NormalTok{b }\OtherTok{:}\NormalTok{ B}\OtherTok{)} \OtherTok{→}\NormalTok{ isContr }\OtherTok{(}\NormalTok{Σ A }\OtherTok{(λ}\NormalTok{ a }\OtherTok{→}\NormalTok{ f a ≡ b}\OtherTok{))}

\NormalTok{    mkInv }\OtherTok{:} \OtherTok{∀} \OtherTok{\{}\NormalTok{ℓ κ}\OtherTok{\}} \OtherTok{\{}\NormalTok{A }\OtherTok{:} \DataTypeTok{Set}\NormalTok{ ℓ}\OtherTok{\}} \OtherTok{\{}\NormalTok{B }\OtherTok{:} \DataTypeTok{Set}\NormalTok{ κ}\OtherTok{\}}
            \OtherTok{→} \OtherTok{(}\NormalTok{f }\OtherTok{:}\NormalTok{ A }\OtherTok{→}\NormalTok{ B}\OtherTok{)} \OtherTok{→}\NormalTok{ isEquiv f }\OtherTok{→}\NormalTok{ B }\OtherTok{→}\NormalTok{ A}
\NormalTok{    mkInv f e b }\OtherTok{=}\NormalTok{ fst }\OtherTok{(}\NormalTok{fst }\OtherTok{(}\NormalTok{e b}\OtherTok{))}

\KeywordTok{open}\NormalTok{ hott}
\end{Highlighting}
\end{Shaded}

The reader familiar with HoTT may note that, so far, we have not
included anything relating to the \emph{univalence axiom} -- arguably
the characteristic axiom of HoTT. In fact, this is by design, as a goal
of the current formalization is to assume only axioms that can be given
straightforward computational interpretations that preserve the property
that every closed term of the ambient type theory evaluates to a
\emph{canonical normal form} (canonicity), so that these axioms give a
\emph{constructive} and \emph{computationally sound} interpretation of
parametricity. While the univalence axiom \emph{can} be given a
computational interpretation compatible with canonicity, as in Cubical
Type Theory \cite{Coquand2022}, doing so is decidedly \emph{not} a straightforward matter.
Moreover, it turns out that the univalence axiom is largely unneeded in
what follows, save for demonstrating admissibility of some additional
axioms which permit more straightforward computational interpretations
that are (conjecturally) compatible with canonicity. I thus shall have
need for univalence only as a metatheoretic assumption. In this setting,
univalence allows us to convert equivalences between types into
\emph{identifications} of those types, which may then be transported
over accordingly.

Having defined some essential structures of the language of HoTT, we may
now proceed to similarly define some essential structures of the
language of HoTT \emph{with cohesive modalities}.

\begin{Shaded}
\begin{Highlighting}[]
\KeywordTok{module}\NormalTok{ cohesion }\KeywordTok{where}
\end{Highlighting}
\end{Shaded}

Of principal importance here is the \(\flat\) modality, which
(intuitively) takes a type \(A\) to its corresponding
\emph{discretization} \(\flat A\), wherein all cohesion between points
has been eliminated. However, in order for this operation to be
well-behaved, \(A\) must not depend upon any variables whose types are
not themselves \emph{discrete}, in this sense. To solve this problem,
the \texttt{-\/-cohesion} flag introduces a new form of variable binding
\texttt{@♭\ x\ :\ X}, which metatheoretically asserts that \(x\) is an
element of the discretization of \(X\), such that \(X\) may only depend
upon variables bound with \texttt{@♭}. In this case, we say that \(x\)
is a \emph{crisp} element of \(X\).

With this notion in hand, we can define the \(\flat\) modality as an
operation on \emph{crisp} types:

\begin{Shaded}
\begin{Highlighting}[]
    \KeywordTok{data}\NormalTok{ ♭ }\OtherTok{\{@}\NormalTok{♭ ℓ }\OtherTok{:}\NormalTok{ Level}\OtherTok{\}} \OtherTok{(@}\NormalTok{♭ A }\OtherTok{:} \DataTypeTok{Set}\NormalTok{ ℓ}\OtherTok{)} \OtherTok{:} \DataTypeTok{Set}\NormalTok{ ℓ }\KeywordTok{where}
\NormalTok{        con }\OtherTok{:} \OtherTok{(@}\NormalTok{♭ x }\OtherTok{:}\NormalTok{ A}\OtherTok{)} \OtherTok{→}\NormalTok{ ♭ A}
\end{Highlighting}
\end{Shaded}

As expected, the \(\flat\) modality is a comonad with the following
counit operation \(\epsilon\):

\begin{Shaded}
\begin{Highlighting}[]
\NormalTok{    ε }\OtherTok{:} \OtherTok{\{@}\NormalTok{♭ l }\OtherTok{:}\NormalTok{ Level}\OtherTok{\}} \OtherTok{\{@}\NormalTok{♭ A }\OtherTok{:} \DataTypeTok{Set}\NormalTok{ l}\OtherTok{\}} \OtherTok{→}\NormalTok{ ♭ A }\OtherTok{→}\NormalTok{ A}
\NormalTok{    ε }\OtherTok{(}\NormalTok{con x}\OtherTok{)} \OtherTok{=}\NormalTok{ x}
\end{Highlighting}
\end{Shaded}

A crisp type is then \emph{discrete} precisely when this map is an
equivalence:

\begin{Shaded}
\begin{Highlighting}[]
\NormalTok{    isDiscrete }\OtherTok{:} \OtherTok{∀} \OtherTok{\{@}\NormalTok{♭ ℓ }\OtherTok{:}\NormalTok{ Level}\OtherTok{\}} \OtherTok{→} \OtherTok{(@}\NormalTok{♭ A }\OtherTok{:} \DataTypeTok{Set}\NormalTok{ ℓ}\OtherTok{)} \OtherTok{→} \DataTypeTok{Set}\NormalTok{ ℓ}
\NormalTok{    isDiscrete }\OtherTok{\{}\NormalTok{ℓ }\OtherTok{=}\NormalTok{ ℓ}\OtherTok{\}}\NormalTok{ A }\OtherTok{=}\NormalTok{ isEquiv }\OtherTok{(}\NormalTok{ε }\OtherTok{\{}\NormalTok{ℓ}\OtherTok{\}} \OtherTok{\{}\NormalTok{A}\OtherTok{\})}

\KeywordTok{open}\NormalTok{ cohesion}
\end{Highlighting}
\end{Shaded}

Beyond such notions of \emph{discreteness}, etc., what more is required for the sake of parametricity is some way of detecting when the elements of
a given type are \emph{related,} or somehow bound together, by the
cohesive structure of that type.

For this purpose, it is useful to take a geometric perspective upon
cohesion, and correspondingly, parametricity. What we are after is
essentially the \emph{shape} of an abstract relation between points, and
an object \(I\) in our cohesive topos \(\mathcal{E}\) (correspondingly,
a type in our type theory) which \emph{classifies} this shape in other
objects (types) in that maps \(I \to A\) correspond to such abstract
relations between points in \(A\). In this case, the \emph{shape} of an
abstract relation may be considered as a \emph{path}, i.e.~two
distinct points which are somehow \emph{connected}. By way of concrete
example, in the topos of reflexive graphs \(\mathbf{RGph}\), the role of
a classifier for this shape is played by the ``walking edge'' graph
\(\bullet \to \bullet\), consisting of two points and a single
non-identity (directed) edge. More generally, using the language of
cohesion, we can capture this notion of an abstract line segment in the
following axiomatic characterization of \(I\):

\begin{quote}
\(I\) is an object of \(\mathcal{E}\) that is \emph{strictly bipointed}
and \emph{weakly connected}.
\end{quote}

Unpacking the terms used in this characterization, we have the
following:

\begin{itemize}
\tightlist
\item
  \emph{Strictly bipointed} means that \(I\) is equipped with a choice
  of two elements \(i_0, i_1 : I\), such that the proposition
  \((i_0 = i_1) \to \bot\) (i.e.~\(i_0 \neq i_1\)) holds in the internal
  language of \(\mathcal{E}\).
\item
  \emph{Weakly connected} means that the unit map
  \(\eta : I \to \smallint I\) is essentially constant, in that it
  factors through a contractible object/type. Intuitively, this says
  that the image of \(I\) in \(\smallint I\) essentially consists of a
  single connected component.
\end{itemize}

Note that the above-given example of the walking edge graph
straightforwardly satisfies both of these requirements, as it consists
of two distinct vertices belonging to a single (weakly) connected
component. I also note in passing that, if the assumption of weak
connectedness is strengthened to \emph{strong connectedness} -- i.e.~the
object/type \(\smallint I\) is itself contractible -- then the existence
of such an object \(I\) as above is equivalent to Lawvere's axiom of
\emph{sufficient cohesion} \cite{LawvereCohesion}. We might therefore refer to the conjunction
of the above conditions as an axiom of \emph{weak sufficient cohesion}
for the ambient \(\infty\)-topos \(\mathcal{E}\).

We can begin to formalize such \emph{weak sufficient cohesion} in Agda
by postulating a type \(I\) with two given elements \(i_0,i_1\):

\begin{Shaded}
\begin{Highlighting}[]
\KeywordTok{postulate}
\NormalTok{    I }\OtherTok{:} \DataTypeTok{Set₀}
\NormalTok{    i0 i1 }\OtherTok{:}\NormalTok{ I}
\end{Highlighting}
\end{Shaded}

We could also, in principle, directly postulate the strict bipointedness of
\(I\), as an axiom having the form \texttt{i0\ ≡\ i1\ →\ ⊥}. However, this
is in fact unnecessary, as this axiom will instead follow from an
equivalent formulation introduced in a following subsection.

On the other hand, we do not yet have the capability to postulate the
axiom of weak connectedness as written above, since we have not yet
formalized the \(\smallint\) modality. We could do so, but again, it is
in fact better for present purposes to rephrase this axiom in an
equivalent form involving only the \(\flat\) modality, which can be done
as follows:

\begin{quote}
A type \(A\) is connected if and only if, for every \emph{discrete} type
\(B\), any function \(A \to B\) is essentially constant, in the sense of
factoring through a contractible type.
\end{quote}

To see that this equivalence holds: in one direction, assume that \(A\)
is weakly connected. Then for any map \(f : A \to B\), by the adjunction
\(\smallint \dashv \flat\) and discreteness of \(B\), there exist maps
\(f_\flat : A \to \flat B\) and \(f^\smallint : \smallint A \to B\),
such that following diagram commutes: \[
\begin{tikzcd}
    {A} & {\smallint A} \\
    {\flat B} & B
    \arrow["{f_\flat}"{description}, from=1-1, to=2-1]
    \arrow["{f^\smallint}"{description}, from=1-2, to=2-2]
    \arrow["{\epsilon}"{description}, from=2-1, to=2-2]
    \arrow["{\eta}"{description}, from=1-1, to=1-2]
    \arrow["f"{description}, from=1-1, to=2-2]
\end{tikzcd}
\] Then since by assumption \(\eta\) factors through a contractible
type, so does \(f\).

In the other direction, assume that every map \(f : A \to B\) is
essentially constant, for every discrete type \(B\). Then in particular,
the map \(\eta : A \to \smallint A\) is essentially constant, since
\(\smallint A\) is discrete (as it lies in the image of the
\emph{discretization} functor \(\Delta\)).

Hence the property of \(I\) being weakly connected can be expressed
purely in terms of its relation to the \emph{discrete types}.
Specifically, if we think of maps \(I \to A\) as abstract
\emph{relations} or \emph{paths} between elements of \(A\), then weak
connectedness of \(I\) equivalently says that \emph{all paths between
points of discrete types are constant}.

In order to conveniently express this property in Agda, it shall
therefore be prudent first to introduce some additional constructs for
ergonomically handling paths, analogous to the definition of path types
in Cubical Type Theory (as in e.g. \cite{CCHM}, \cite{ABCHFL}).

\subsection{Path Types}\label{path-types}

In principle, given \(a,b : A\), we could define the type of paths from
\(a\) to \(b\) in \(A\) as the type
\(\Sigma f : I \to A . (f ~ i_0 = a) \times (f ~ i_1 = b)\). However,
experience with such a naïve formalization shows that it incurs a high
number of laborious transportations along equalities that should be easy
enough to infer automatically. Hence I instead follow the approach taken
by Cubical Type Theory and related systems, and give an explicit
axiomatization for \emph{path types}, with corresponding rewrite rules
to apply the associated equalities automatically:

\begin{Shaded}
\begin{Highlighting}[]
\KeywordTok{postulate}
\NormalTok{    Path }\OtherTok{:} \OtherTok{∀} \OtherTok{\{}\NormalTok{ℓ}\OtherTok{\}} \OtherTok{(}\NormalTok{A }\OtherTok{:}\NormalTok{ I }\OtherTok{→} \DataTypeTok{Set}\NormalTok{ ℓ}\OtherTok{)} \OtherTok{(}\NormalTok{a0 }\OtherTok{:}\NormalTok{ A i0}\OtherTok{)} \OtherTok{(}\NormalTok{a1 }\OtherTok{:}\NormalTok{ A i1}\OtherTok{)} \OtherTok{→} \DataTypeTok{Set}\NormalTok{ ℓ}
\end{Highlighting}
\end{Shaded}

The introduction rule for path types corresponds to \emph{function
abstraction}

\begin{Shaded}
\begin{Highlighting}[]
\NormalTok{    pabs }\OtherTok{:} \OtherTok{∀} \OtherTok{\{}\NormalTok{ℓ}\OtherTok{\}} \OtherTok{\{}\NormalTok{A }\OtherTok{:}\NormalTok{ I }\OtherTok{→} \DataTypeTok{Set}\NormalTok{ ℓ}\OtherTok{\}} 
           \OtherTok{→} \OtherTok{(}\NormalTok{f }\OtherTok{:} \OtherTok{(}\NormalTok{i }\OtherTok{:}\NormalTok{ I}\OtherTok{)} \OtherTok{→}\NormalTok{ A i}\OtherTok{)} \OtherTok{→}\NormalTok{ Path A }\OtherTok{(}\NormalTok{f i0}\OtherTok{)} \OtherTok{(}\NormalTok{f i1}\OtherTok{)}
\end{Highlighting}
\end{Shaded}

and likewise, the elimination rule corresponds to \emph{function
application}.

\begin{Shaded}
\begin{Highlighting}[]
\NormalTok{    papp }\OtherTok{:} \OtherTok{∀} \OtherTok{\{}\NormalTok{ℓ}\OtherTok{\}} \OtherTok{\{}\NormalTok{A }\OtherTok{:}\NormalTok{ I }\OtherTok{→} \DataTypeTok{Set}\NormalTok{ ℓ}\OtherTok{\}} \OtherTok{\{}\NormalTok{a0 }\OtherTok{:}\NormalTok{ A i0}\OtherTok{\}} \OtherTok{\{}\NormalTok{a1 }\OtherTok{:}\NormalTok{ A i1}\OtherTok{\}}
           \OtherTok{→}\NormalTok{ Path A a0 a1 }\OtherTok{→} \OtherTok{(}\NormalTok{i }\OtherTok{:}\NormalTok{ I}\OtherTok{)} \OtherTok{→}\NormalTok{ A i}
\end{Highlighting}
\end{Shaded}

We may then postulate the usual \(\beta\)-law as an identity for this
type, along with special identities for application to \(i_0\) and
\(i_1\). All of these are made into rewrite rules, allowing Agda to apply
them automatically, and thus obviating the need for excessive use of
transport:\footnote{We could additionally postulate an \(\eta\)-law for
  path types, analogous to the usual \(\eta\)-law for function types;
  however, this is unnecessary for what follows, and so I omit this
  assumption.}

\begin{Shaded}
\begin{Highlighting}[]
\NormalTok{    pβ }\OtherTok{:} \OtherTok{∀} \OtherTok{\{}\NormalTok{ℓ}\OtherTok{\}} \OtherTok{\{}\NormalTok{A }\OtherTok{:}\NormalTok{ I }\OtherTok{→} \DataTypeTok{Set}\NormalTok{ ℓ}\OtherTok{\}} \OtherTok{(}\NormalTok{f }\OtherTok{:} \OtherTok{(}\NormalTok{i }\OtherTok{:}\NormalTok{ I}\OtherTok{)} \OtherTok{→}\NormalTok{ A i}\OtherTok{)} 
           \OtherTok{→} \OtherTok{(}\NormalTok{i }\OtherTok{:}\NormalTok{ I}\OtherTok{)} \OtherTok{→}\NormalTok{ papp }\OtherTok{(}\NormalTok{pabs f}\OtherTok{)}\NormalTok{ i ≡ f i}
    \PreprocessorTok{\{{-}\# REWRITE pβ \#{-}\}}
\NormalTok{    papp0 }\OtherTok{:} \OtherTok{∀} \OtherTok{\{}\NormalTok{ℓ}\OtherTok{\}} \OtherTok{\{}\NormalTok{A }\OtherTok{:}\NormalTok{ I }\OtherTok{→} \DataTypeTok{Set}\NormalTok{ ℓ}\OtherTok{\}} \OtherTok{\{}\NormalTok{a0 }\OtherTok{:}\NormalTok{ A i0}\OtherTok{\}} \OtherTok{\{}\NormalTok{a1 }\OtherTok{:}\NormalTok{ A i1}\OtherTok{\}} 
            \OtherTok{→} \OtherTok{(}\NormalTok{p }\OtherTok{:}\NormalTok{ Path A a0 a1}\OtherTok{)} \OtherTok{→}\NormalTok{ papp p i0 ≡ a0}
    \PreprocessorTok{\{{-}\# REWRITE papp0 \#{-}\}}
\NormalTok{    papp1 }\OtherTok{:} \OtherTok{∀} \OtherTok{\{}\NormalTok{ℓ}\OtherTok{\}} \OtherTok{\{}\NormalTok{A }\OtherTok{:}\NormalTok{ I }\OtherTok{→} \DataTypeTok{Set}\NormalTok{ ℓ}\OtherTok{\}} \OtherTok{\{}\NormalTok{a0 }\OtherTok{:}\NormalTok{ A i0}\OtherTok{\}} \OtherTok{\{}\NormalTok{a1 }\OtherTok{:}\NormalTok{ A i1}\OtherTok{\}} 
            \OtherTok{→} \OtherTok{(}\NormalTok{p }\OtherTok{:}\NormalTok{ Path A a0 a1}\OtherTok{)} \OtherTok{→}\NormalTok{ papp p i1 ≡ a1}
    \PreprocessorTok{\{{-}\# REWRITE papp1 \#{-}\}}
\end{Highlighting}
\end{Shaded}

With this formalization of path types in hand, we can straightforwardly
formalize the equivalent formulation of weak connectedness of \(I\)
given above. For this purpose, we first define the map \texttt{idToPath}
that takes an identification \(a ≡ b\) to a path from \(a\) to \(b\):

\begin{Shaded}
\begin{Highlighting}[]
\NormalTok{idToPath }\OtherTok{:} \OtherTok{∀} \OtherTok{\{}\NormalTok{ℓ}\OtherTok{\}} \OtherTok{\{}\NormalTok{A }\OtherTok{:} \DataTypeTok{Set}\NormalTok{ ℓ}\OtherTok{\}} \OtherTok{\{}\NormalTok{a b }\OtherTok{:}\NormalTok{ A}\OtherTok{\}}
           \OtherTok{→}\NormalTok{ a ≡ b }\OtherTok{→}\NormalTok{ Path }\OtherTok{(λ} \OtherTok{\_} \OtherTok{→}\NormalTok{ A}\OtherTok{)}\NormalTok{ a b}
\NormalTok{idToPath }\OtherTok{\{}\NormalTok{a }\OtherTok{=}\NormalTok{ a}\OtherTok{\}}\NormalTok{ refl }\OtherTok{=}\NormalTok{ pabs }\OtherTok{(λ} \OtherTok{\_} \OtherTok{→}\NormalTok{ a}\OtherTok{)}
\end{Highlighting}
\end{Shaded}

A type \(A\) is \emph{path-discrete} if for all \(a,b : A\) the map
\texttt{idToPath} is an equivalence:

\begin{Shaded}
\begin{Highlighting}[]
\NormalTok{isPathDiscrete }\OtherTok{:} \OtherTok{∀} \OtherTok{\{}\NormalTok{ℓ}\OtherTok{\}} \OtherTok{(}\NormalTok{A }\OtherTok{:} \DataTypeTok{Set}\NormalTok{ ℓ}\OtherTok{)} \OtherTok{→} \DataTypeTok{Set}\NormalTok{ ℓ}
\NormalTok{isPathDiscrete }\OtherTok{\{}\NormalTok{ℓ }\OtherTok{=}\NormalTok{ ℓ}\OtherTok{\}}\NormalTok{ A }\OtherTok{=} 
    \OtherTok{\{}\NormalTok{a b }\OtherTok{:}\NormalTok{ A}\OtherTok{\}} \OtherTok{→}\NormalTok{ isEquiv }\OtherTok{(}\NormalTok{idToPath }\OtherTok{\{}\NormalTok{ℓ}\OtherTok{\}} \OtherTok{\{}\NormalTok{A}\OtherTok{\}} \OtherTok{\{}\NormalTok{a}\OtherTok{\}} \OtherTok{\{}\NormalTok{b}\OtherTok{\})}
\end{Highlighting}
\end{Shaded}

We then postulate the following axioms:

\begin{Shaded}
\begin{Highlighting}[]
\KeywordTok{postulate}
\NormalTok{    pathConst1 }\OtherTok{:} \OtherTok{∀} \OtherTok{\{@}\NormalTok{♭ ℓ }\OtherTok{:}\NormalTok{ Level}\OtherTok{\}} \OtherTok{\{@}\NormalTok{♭ A }\OtherTok{:} \DataTypeTok{Set}\NormalTok{ ℓ}\OtherTok{\}} \OtherTok{\{}\NormalTok{a b }\OtherTok{:}\NormalTok{ A}\OtherTok{\}}
                   \OtherTok{→}\NormalTok{ isDiscrete A }\OtherTok{→} \OtherTok{(}\NormalTok{e }\OtherTok{:}\NormalTok{ Path }\OtherTok{(λ} \OtherTok{\_} \OtherTok{→}\NormalTok{ A}\OtherTok{)}\NormalTok{ a b}\OtherTok{)}
                   \OtherTok{→}\NormalTok{ Σ }\OtherTok{(}\NormalTok{a ≡ b}\OtherTok{)} \OtherTok{(λ}\NormalTok{ p }\OtherTok{→}\NormalTok{ idToPath p ≡ e}\OtherTok{)}
\NormalTok{    pathConst2 }\OtherTok{:} \OtherTok{∀} \OtherTok{\{@}\NormalTok{♭ ℓ }\OtherTok{:}\NormalTok{ Level}\OtherTok{\}} \OtherTok{\{@}\NormalTok{♭ A }\OtherTok{:} \DataTypeTok{Set}\NormalTok{ ℓ}\OtherTok{\}} \OtherTok{\{}\NormalTok{a b }\OtherTok{:}\NormalTok{ A}\OtherTok{\}}
                   \OtherTok{→} \OtherTok{(}\NormalTok{dA }\OtherTok{:}\NormalTok{ isDiscrete A}\OtherTok{)} \OtherTok{→} \OtherTok{(}\NormalTok{e }\OtherTok{:}\NormalTok{ Path }\OtherTok{(λ} \OtherTok{\_} \OtherTok{→}\NormalTok{ A}\OtherTok{)}\NormalTok{ a b}\OtherTok{)}
                   \OtherTok{→} \OtherTok{(}\NormalTok{q }\OtherTok{:}\NormalTok{ a ≡ b}\OtherTok{)} \OtherTok{→} \OtherTok{(}\NormalTok{r }\OtherTok{:}\NormalTok{ idToPath q ≡ e}\OtherTok{)}
                   \OtherTok{→}\NormalTok{ pathConst1 dA e ≡ }\OtherTok{(}\NormalTok{q , r}\OtherTok{)}
\end{Highlighting}
\end{Shaded}

which together imply that, if \(A\) is discrete, then it is
\emph{path-discrete}:

\begin{Shaded}
\begin{Highlighting}[]
\NormalTok{isDisc→isPDisc }\OtherTok{:} \OtherTok{∀} \OtherTok{\{@}\NormalTok{♭ ℓ }\OtherTok{:}\NormalTok{ Level}\OtherTok{\}} \OtherTok{\{@}\NormalTok{♭ A }\OtherTok{:} \DataTypeTok{Set}\NormalTok{ ℓ}\OtherTok{\}}
                 \OtherTok{→}\NormalTok{ isDiscrete A }\OtherTok{→}\NormalTok{ isPathDiscrete A}
\NormalTok{isDisc→isPDisc dA e }\OtherTok{=} 
    \OtherTok{(}\NormalTok{pathConst1 dA e , }\OtherTok{λ} \OtherTok{(}\NormalTok{p , q}\OtherTok{)} \OtherTok{→}\NormalTok{ pathConst2 dA e p q}\OtherTok{)}
\end{Highlighting}
\end{Shaded}

As it stands, we have not yet given a procedure for evaluating the
axioms \texttt{pathConst1} and \texttt{pathConst2} when they are applied
to canonical forms, which means that computation on these terms will
generally get stuck and thus violate canonicity. Toward rectifying this,
I prove a key identity regarding these axioms, add a further postulate
asserting that this identity is equal to \texttt{refl}, and convert both
of these to rewrite rules:

\begin{Shaded}
\begin{Highlighting}[]
\NormalTok{rwPathConst1 }\OtherTok{:} \OtherTok{∀} \OtherTok{\{@}\NormalTok{♭ ℓ }\OtherTok{:}\NormalTok{ Level}\OtherTok{\}} \OtherTok{\{@}\NormalTok{♭ A }\OtherTok{:} \DataTypeTok{Set}\NormalTok{ ℓ}\OtherTok{\}} \OtherTok{\{}\NormalTok{a }\OtherTok{:}\NormalTok{ A}\OtherTok{\}} \OtherTok{→} \OtherTok{(}\NormalTok{dA }\OtherTok{:}\NormalTok{ isDiscrete A}\OtherTok{)} 
               \OtherTok{→}\NormalTok{ pathConst1 dA }\OtherTok{(}\NormalTok{pabs }\OtherTok{(λ} \OtherTok{\_} \OtherTok{→}\NormalTok{ a}\OtherTok{))}\NormalTok{ ≡ }\OtherTok{(}\NormalTok{refl , refl}\OtherTok{)}
\NormalTok{rwPathConst1 }\OtherTok{\{}\NormalTok{a }\OtherTok{=}\NormalTok{ a}\OtherTok{\}}\NormalTok{ dA }\OtherTok{=}\NormalTok{ pathConst2 dA }\OtherTok{(}\NormalTok{pabs }\OtherTok{(λ} \OtherTok{\_} \OtherTok{→}\NormalTok{ a}\OtherTok{))}\NormalTok{ refl refl}
\PreprocessorTok{\{{-}\# REWRITE rwPathConst1 \#{-}\}}

\KeywordTok{postulate}
\NormalTok{    rwPathConst2 }\OtherTok{:} \OtherTok{∀} \OtherTok{\{@}\NormalTok{♭ ℓ }\OtherTok{:}\NormalTok{ Level}\OtherTok{\}} \OtherTok{\{@}\NormalTok{♭ A }\OtherTok{:} \DataTypeTok{Set}\NormalTok{ ℓ}\OtherTok{\}} \OtherTok{\{}\NormalTok{a }\OtherTok{:}\NormalTok{ A}\OtherTok{\}} \OtherTok{→} \OtherTok{(}\NormalTok{dA }\OtherTok{:}\NormalTok{ isDiscrete A}\OtherTok{)}
                   \OtherTok{→}\NormalTok{ pathConst2 dA }\OtherTok{(}\NormalTok{pabs }\OtherTok{(λ} \OtherTok{\_} \OtherTok{→}\NormalTok{ a}\OtherTok{))}\NormalTok{ refl refl ≡ refl}
    \PreprocessorTok{\{{-}\# REWRITE rwPathConst2 \#{-}\}}
\end{Highlighting}
\end{Shaded}

Although a full proof of canonicity is beyond the scope of this paper, I
conjecture that adding these rules suffices to preserve canonicity, and
I verify a few concrete cases of this conjecture later in the paper.

So much for the (weak) connectedness of \(I\); let us now turn our
attention to the other property we had previously stipulated of \(I\),
namely its \emph{strict bipointedness}. As mentioned previously, we
could simply postulate this stipulation directly as an axiom -- however,
for the purpose of proving parametricity theorems, a more prudent
strategy is to instead formalize a class of \(I\)-indexed type families,
whose computational behavior follows from this assumption (and which, in
turn, implies it). Because these type families essentially correspond to
the \emph{graphs} of predicates and relations on arbitrary types, I
refer to them as \emph{graph types}.

\subsection{Graph Types}\label{graph-types}

For present purposes, we need only concern ourselves with the simplest
class of graph types: \emph{unary graph types}, which, as the name would
imply, correspond to graphs of unary predicates. Given a type \(A\), a
type family \(B : A \to \mathsf{Type}\), and an element \(i : I\), the
\emph{graph type} \(\mathsf{Gph}^1 ~ i ~ A ~ B\) is defined to be equal
to \(A\) when \(i\) is \(i_0\), and equivalent to \(\Sigma x : A . B x\)
when \(i\) is \(i_1\). Intuitively, an element of
\(\mathsf{Gph}^1 ~ i ~ A ~ B\) is a dependent pair whose second element
\emph{only exists when \(i\) is equal to \(i_1\)}. We may formalize this
in Agda as follows, by postulating a rewrite rule that evaluates
\(\mathsf{Gph}^1 ~ i_0 ~ A ~ B\) to \(A\):

\begin{Shaded}
\begin{Highlighting}[]
\KeywordTok{postulate}
\NormalTok{    Gph1 }\OtherTok{:} \OtherTok{∀} \OtherTok{\{}\NormalTok{ℓ}\OtherTok{\}} \OtherTok{(}\NormalTok{i }\OtherTok{:}\NormalTok{ I}\OtherTok{)} \OtherTok{(}\NormalTok{A }\OtherTok{:} \DataTypeTok{Set}\NormalTok{ ℓ}\OtherTok{)} \OtherTok{(}\NormalTok{B }\OtherTok{:}\NormalTok{ A }\OtherTok{→} \DataTypeTok{Set}\NormalTok{ ℓ}\OtherTok{)} \OtherTok{→} \DataTypeTok{Set} \OtherTok{(}\NormalTok{ℓ}\OtherTok{)}

\NormalTok{    g1rw0 }\OtherTok{:} \OtherTok{∀} \OtherTok{\{}\NormalTok{ℓ}\OtherTok{\}} \OtherTok{(}\NormalTok{A }\OtherTok{:} \DataTypeTok{Set}\NormalTok{ ℓ}\OtherTok{)} \OtherTok{(}\NormalTok{B }\OtherTok{:}\NormalTok{ A }\OtherTok{→} \DataTypeTok{Set}\NormalTok{ ℓ}\OtherTok{)} \OtherTok{→}\NormalTok{ Gph1 i0 A B ≡ A}
    \PreprocessorTok{\{{-}\# REWRITE g1rw0 \#{-}\}}
\end{Highlighting}
\end{Shaded}

We then have the following introduction rule for elements of
\(\mathsf{Gph}^1 ~ i ~ A ~ B\), which are pairs where the second element
of the pair only exists under the assumption that \(i = i_1\). When
\(i = i_0\) instead, the pair collapses to its first element:

\begin{Shaded}
\begin{Highlighting}[]
\NormalTok{    g1pair }\OtherTok{:} \OtherTok{∀} \OtherTok{\{}\NormalTok{ℓ}\OtherTok{\}} \OtherTok{\{}\NormalTok{A }\OtherTok{:} \DataTypeTok{Set}\NormalTok{ ℓ}\OtherTok{\}} \OtherTok{\{}\NormalTok{B }\OtherTok{:}\NormalTok{ A }\OtherTok{→} \DataTypeTok{Set}\NormalTok{ ℓ}\OtherTok{\}} \OtherTok{(}\NormalTok{i }\OtherTok{:}\NormalTok{ I}\OtherTok{)}
             \OtherTok{→} \OtherTok{(}\NormalTok{a }\OtherTok{:}\NormalTok{ A}\OtherTok{)} \OtherTok{→} \OtherTok{(}\NormalTok{b }\OtherTok{:} \OtherTok{(}\NormalTok{i ≡ i1}\OtherTok{)} \OtherTok{→}\NormalTok{ B a}\OtherTok{)} \OtherTok{→}\NormalTok{ Gph1 i A B}

\NormalTok{    g1pair0 }\OtherTok{:} \OtherTok{∀} \OtherTok{\{}\NormalTok{ℓ}\OtherTok{\}} \OtherTok{\{}\NormalTok{A }\OtherTok{:} \DataTypeTok{Set}\NormalTok{ ℓ}\OtherTok{\}} \OtherTok{\{}\NormalTok{B }\OtherTok{:}\NormalTok{ A }\OtherTok{→} \DataTypeTok{Set}\NormalTok{ ℓ}\OtherTok{\}}
              \OtherTok{→} \OtherTok{(}\NormalTok{a }\OtherTok{:}\NormalTok{ A}\OtherTok{)} \OtherTok{→} \OtherTok{(}\NormalTok{b }\OtherTok{:} \OtherTok{(}\NormalTok{i0 ≡ i1}\OtherTok{)} \OtherTok{→}\NormalTok{ B a}\OtherTok{)}
              \OtherTok{→}\NormalTok{ g1pair }\OtherTok{\{}\NormalTok{B }\OtherTok{=}\NormalTok{ B}\OtherTok{\}}\NormalTok{ i0 a b ≡ a}
    \PreprocessorTok{\{{-}\# REWRITE g1pair0 \#{-}\}}
\end{Highlighting}
\end{Shaded}

The first projection from such a pair may then be taken no matter what
\(i\) is, and reduces to the identity function when \(i\) is \(i_0\):

\begin{Shaded}
\begin{Highlighting}[]
\NormalTok{    g1fst }\OtherTok{:} \OtherTok{∀} \OtherTok{\{}\NormalTok{ℓ}\OtherTok{\}} \OtherTok{\{}\NormalTok{A }\OtherTok{:} \DataTypeTok{Set}\NormalTok{ ℓ}\OtherTok{\}} \OtherTok{\{}\NormalTok{B }\OtherTok{:}\NormalTok{ A }\OtherTok{→} \DataTypeTok{Set}\NormalTok{ ℓ}\OtherTok{\}} \OtherTok{(}\NormalTok{i }\OtherTok{:}\NormalTok{ I}\OtherTok{)}
            \OtherTok{→} \OtherTok{(}\NormalTok{g }\OtherTok{:}\NormalTok{ Gph1 i A B}\OtherTok{)} \OtherTok{→}\NormalTok{ A}
    
\NormalTok{    g1beta1 }\OtherTok{:} \OtherTok{∀} \OtherTok{\{}\NormalTok{ℓ}\OtherTok{\}} \OtherTok{\{}\NormalTok{A }\OtherTok{:} \DataTypeTok{Set}\NormalTok{ ℓ}\OtherTok{\}} \OtherTok{\{}\NormalTok{B }\OtherTok{:}\NormalTok{ A }\OtherTok{→} \DataTypeTok{Set}\NormalTok{ ℓ}\OtherTok{\}} \OtherTok{(}\NormalTok{i }\OtherTok{:}\NormalTok{ I}\OtherTok{)}
              \OtherTok{→} \OtherTok{(}\NormalTok{a }\OtherTok{:}\NormalTok{ A}\OtherTok{)} \OtherTok{→} \OtherTok{(}\NormalTok{b }\OtherTok{:} \OtherTok{(}\NormalTok{i ≡ i1}\OtherTok{)} \OtherTok{→}\NormalTok{ B a}\OtherTok{)}
              \OtherTok{→}\NormalTok{ g1fst i }\OtherTok{(}\NormalTok{g1pair }\OtherTok{\{}\NormalTok{B }\OtherTok{=}\NormalTok{ B}\OtherTok{\}}\NormalTok{ i a b}\OtherTok{)}\NormalTok{ ≡ a}
    \PreprocessorTok{\{{-}\# REWRITE g1beta1 \#{-}\}}
    
\NormalTok{    g1fst0 }\OtherTok{:} \OtherTok{∀} \OtherTok{\{}\NormalTok{ℓ}\OtherTok{\}} \OtherTok{\{}\NormalTok{A }\OtherTok{:} \DataTypeTok{Set}\NormalTok{ ℓ}\OtherTok{\}} \OtherTok{\{}\NormalTok{B }\OtherTok{:}\NormalTok{ A }\OtherTok{→} \DataTypeTok{Set}\NormalTok{ ℓ}\OtherTok{\}}
             \OtherTok{→} \OtherTok{(}\NormalTok{g }\OtherTok{:}\NormalTok{ Gph1 i0 A B}\OtherTok{)} \OtherTok{→}\NormalTok{ g1fst }\OtherTok{\{}\NormalTok{B }\OtherTok{=}\NormalTok{ B}\OtherTok{\}}\NormalTok{ i0 g ≡ g}
    \PreprocessorTok{\{{-}\# REWRITE g1fst0 \#{-}\}}
\end{Highlighting}
\end{Shaded}

The second projection, meanwhile, may only be taken when \(i\) is equal
to \(i_1\):

\begin{Shaded}
\begin{Highlighting}[]
\NormalTok{    g1snd }\OtherTok{:} \OtherTok{∀} \OtherTok{\{}\NormalTok{ℓ}\OtherTok{\}} \OtherTok{\{}\NormalTok{A }\OtherTok{:} \DataTypeTok{Set}\NormalTok{ ℓ}\OtherTok{\}} \OtherTok{\{}\NormalTok{B }\OtherTok{:}\NormalTok{ A }\OtherTok{→} \DataTypeTok{Set}\NormalTok{ ℓ}\OtherTok{\}}
            \OtherTok{→} \OtherTok{(}\NormalTok{g }\OtherTok{:}\NormalTok{ Gph1 i1 A B}\OtherTok{)} \OtherTok{→}\NormalTok{ B }\OtherTok{(}\NormalTok{g1fst i1 g}\OtherTok{)}
    
\NormalTok{    g1beta2 }\OtherTok{:} \OtherTok{∀} \OtherTok{\{}\NormalTok{ℓ}\OtherTok{\}} \OtherTok{\{}\NormalTok{A }\OtherTok{:} \DataTypeTok{Set}\NormalTok{ ℓ}\OtherTok{\}} \OtherTok{\{}\NormalTok{B }\OtherTok{:}\NormalTok{ A }\OtherTok{→} \DataTypeTok{Set}\NormalTok{ ℓ}\OtherTok{\}}
              \OtherTok{→} \OtherTok{(}\NormalTok{a }\OtherTok{:}\NormalTok{ A}\OtherTok{)} \OtherTok{→} \OtherTok{(}\NormalTok{b }\OtherTok{:} \OtherTok{(}\NormalTok{i1 ≡ i1}\OtherTok{)} \OtherTok{→}\NormalTok{ B a}\OtherTok{)}
              \OtherTok{→}\NormalTok{ g1snd }\OtherTok{(}\NormalTok{g1pair }\OtherTok{\{}\NormalTok{B }\OtherTok{=}\NormalTok{ B}\OtherTok{\}}\NormalTok{ i1 a b}\OtherTok{)}\NormalTok{ ≡ b refl}
    \PreprocessorTok{\{{-}\# REWRITE g1beta2 \#{-}\}}
\end{Highlighting}
\end{Shaded}

It is straightforward to see that the inclusion of graph types makes
strict bipointedness of the interval provable, as follows:

\begin{Shaded}
\begin{Highlighting}[]
\NormalTok{strBpt }\OtherTok{:} \OtherTok{(}\NormalTok{i0 ≡ i1}\OtherTok{)} \OtherTok{→}\NormalTok{ ⊥}
\NormalTok{strBpt p }\OtherTok{=}\NormalTok{ g1snd }\OtherTok{(}\NormalTok{transp }\OtherTok{(λ}\NormalTok{ i }\OtherTok{→}\NormalTok{ Gph1 i ⊤ }\OtherTok{(λ} \OtherTok{\_} \OtherTok{→}\NormalTok{ ⊥}\OtherTok{))}\NormalTok{ p tt}\OtherTok{)}
\end{Highlighting}
\end{Shaded}

And in fact, the converse holds under the assumption of univalence.
Specifically, in the presence of univalence and the assumption of strict
bipointedness for \(I\), the type \(\mathsf{Gph}^1 ~ i ~ A ~ B\) may be
regarded as a computationally convenient shorthand for the type
\(\Sigma x : A . (i = i_1) \to B x\), in much the same way as the type
\(\mathsf{Path} ~ A ~ a_0 ~ a_1\) serves as shorthand for the type
\(\Sigma f : (\Pi i : I . A i) . (f ~ i_0 = a_0) \times (f ~ i_1 = a_1)\).
This fact is due to the following equivalence
\[\begin{array}{rl} &\Sigma x : A . (i_0 = i_1) \to B x\\ \simeq & \Sigma x : A . \bot \to B x\\ \simeq & \Sigma x : A . \top \\ \simeq & A \end{array}\]
which is given by the map
\(\mathsf{fst} : (\Sigma x : A . (i_0 = i_1) \to B x) \to A\) and which,
under univalence, becomes an identity between its domain and codomain,
thereby justifying the use of this and associated identities as rewrite
rules which, conjecturally, are fully compatible with canonictiy.

A few additional theorems, concerning identities between elements of
graph types, are as follows, the latter of which I make into a rewrite
rule for convenience:

\begin{Shaded}
\begin{Highlighting}[]
\NormalTok{apg1pair }\OtherTok{:} \OtherTok{∀} \OtherTok{\{}\NormalTok{ℓ}\OtherTok{\}} \OtherTok{\{}\NormalTok{A }\OtherTok{:} \DataTypeTok{Set}\NormalTok{ ℓ}\OtherTok{\}} \OtherTok{\{}\NormalTok{B }\OtherTok{:}\NormalTok{ A }\OtherTok{→} \DataTypeTok{Set}\NormalTok{ ℓ}\OtherTok{\}}
           \OtherTok{→} \OtherTok{\{}\NormalTok{a b }\OtherTok{:}\NormalTok{ A}\OtherTok{\}} \OtherTok{\{}\NormalTok{aB }\OtherTok{:}\NormalTok{ B a}\OtherTok{\}} \OtherTok{\{}\NormalTok{bB }\OtherTok{:}\NormalTok{ B b}\OtherTok{\}} 
           \OtherTok{→} \OtherTok{(}\NormalTok{p }\OtherTok{:}\NormalTok{ a ≡ b}\OtherTok{)} \OtherTok{→}\NormalTok{ aB ≡ transp⁻¹ B p bB }
           \OtherTok{→} \OtherTok{(}\NormalTok{i }\OtherTok{:}\NormalTok{ I}\OtherTok{)} \OtherTok{→}\NormalTok{ g1pair i a }\OtherTok{(λ} \OtherTok{\_} \OtherTok{→}\NormalTok{ aB}\OtherTok{)}\NormalTok{ ≡ g1pair i b }\OtherTok{(λ} \OtherTok{\_} \OtherTok{→}\NormalTok{ bB}\OtherTok{)}
\NormalTok{apg1pair refl refl i }\OtherTok{=}\NormalTok{ refl}

\NormalTok{apg1pair0 }\OtherTok{:} \OtherTok{∀} \OtherTok{\{}\NormalTok{ℓ}\OtherTok{\}} \OtherTok{\{}\NormalTok{A }\OtherTok{:} \DataTypeTok{Set}\NormalTok{ ℓ}\OtherTok{\}} \OtherTok{\{}\NormalTok{B }\OtherTok{:}\NormalTok{ A }\OtherTok{→} \DataTypeTok{Set}\NormalTok{ ℓ}\OtherTok{\}}
            \OtherTok{→} \OtherTok{\{}\NormalTok{a b }\OtherTok{:}\NormalTok{ A}\OtherTok{\}} \OtherTok{\{}\NormalTok{aB }\OtherTok{:}\NormalTok{ B a}\OtherTok{\}} \OtherTok{\{}\NormalTok{bB }\OtherTok{:}\NormalTok{ B b}\OtherTok{\}}
            \OtherTok{→} \OtherTok{(}\NormalTok{p }\OtherTok{:}\NormalTok{ a ≡ b}\OtherTok{)} \OtherTok{→} \OtherTok{(}\NormalTok{q }\OtherTok{:}\NormalTok{ aB ≡ transp⁻¹ B p bB}\OtherTok{)}
            \OtherTok{→}\NormalTok{ apg1pair p q i0 ≡ p}
\NormalTok{apg1pair0 refl refl }\OtherTok{=}\NormalTok{ refl}
\PreprocessorTok{\{{-}\# REWRITE apg1pair0 \#{-}\}}
\end{Highlighting}
\end{Shaded}

In principle, we could continue in this manner, and define graph types
for relations of arities greater than 1 as well. However, unary graph
types are sufficient for what follows. Having thus
given appropriate axioms (and corresponding computation rules) to
capture the desiderata that \(I\) be strictly bipointed and weakly
connected, we are now in a position to prove some classical
parametricity theorems using this structure.

\subsection{Parametricity via Sufficient
Cohesion}\label{parametricity-via-sufficient-cohesion}

I begin this section with an old chestnut of parametricity theorems -- a
proof that any \emph{polymorphic function} of type
\texttt{(X\ :\ Set)\ →\ X\ →\ X} must be equivalent to the polymorphic
identity function \texttt{λ\ X\ →\ λ\ x\ →\ x}.

\begin{Shaded}
\begin{Highlighting}[]
\NormalTok{PolyId }\OtherTok{:} \OtherTok{(}\NormalTok{ℓ }\OtherTok{:}\NormalTok{ Level}\OtherTok{)} \OtherTok{→} \DataTypeTok{Set} \OtherTok{(}\NormalTok{lsuc ℓ}\OtherTok{)}
\NormalTok{PolyId ℓ }\OtherTok{=} \OtherTok{(}\NormalTok{X }\OtherTok{:} \DataTypeTok{Set}\NormalTok{ ℓ}\OtherTok{)} \OtherTok{→}\NormalTok{ X }\OtherTok{→}\NormalTok{ X}
\end{Highlighting}
\end{Shaded}

Before proceeding with this proof, however, it will be prudent to
consider the \emph{meaning} of this theorem in the context of the
cohesive type theory we have so-far developed. Specifically, I wish to
ask: over which types should the type variable \texttt{X} in the type
\texttt{(X\ :\ Set)\ →\ X\ →\ X} be considered as ranging in the
statement of this theorem? Although it is tempting to think that the answer to this question should
be ``all types'' (or as close to this as one can get predicatively), if
one considers the relation between our cohesive setup and Reynolds'
original setup of parametricity, a subtler picture emerges. A type, in
our framework, corresponds not to a type in the object language of
e.g.~bare sets, but rather to an object of the cohesive topos used to
interpret the parametric structure of this object language, e.g.~the
category of reflexive graphs. In this sense, we should expect the
parametricity result for the type \texttt{(X\ :\ Set)\ →\ X\ →\ X} to
generally hold only for those types corresponding to those in the object
language, i.e.~the \emph{discrete types}. Indeed, the discrete types by
construction are those which cannot distinguish elements of other types
belonging to the same connected component, which intuitively corresponds
to the fundamental idea of parametricity -- that functions defined over
these types must behave essentially the same for related inputs.

However, we cannot state this formulation of the theorem directly, since
it would require us to bind \(X\) as \texttt{@♭\ X\ :\ Set}, which would
kill all of the cohesive structure on \texttt{Set} and pose no
restriction on the functions inhabiting this type. The solution, in this
case, is to restrict the range of \texttt{X} to types which are
\emph{path-discrete}, since this requirement can be stated even for
\texttt{X} not crisp.

To prove the desired parametricity theorem for the type \texttt{PolyId}
as above, then, we first prove a \emph{substitution lemma} as follows:

\begin{quote}
For any function \texttt{α\ :\ PolyIdA} and any path-discrete type
\texttt{A} with \texttt{a\ :\ A} and a type family
\texttt{B\ :\ A\ →\ Set}, there is a function
\texttt{B\ a\ →\ B(α\ A\ a)}
\end{quote}

\begin{Shaded}
\begin{Highlighting}[]
\KeywordTok{module}\NormalTok{ paramId }\OtherTok{\{}\NormalTok{ℓ}\OtherTok{\}} \OtherTok{(}\NormalTok{A }\OtherTok{:} \DataTypeTok{Set}\NormalTok{ ℓ}\OtherTok{)} \OtherTok{(}\NormalTok{pdA }\OtherTok{:}\NormalTok{ isPathDiscrete A}\OtherTok{)} \OtherTok{(}\NormalTok{B }\OtherTok{:}\NormalTok{ A }\OtherTok{→} \DataTypeTok{Set}\NormalTok{ ℓ}\OtherTok{)} 
                   \OtherTok{(}\NormalTok{a }\OtherTok{:}\NormalTok{ A}\OtherTok{)} \OtherTok{(}\NormalTok{b }\OtherTok{:}\NormalTok{ B a}\OtherTok{)} \OtherTok{(}\NormalTok{α }\OtherTok{:}\NormalTok{ PolyId ℓ}\OtherTok{)} \KeywordTok{where}
\end{Highlighting}
\end{Shaded}

The key step in the proof of this lemma is to construct a ``dependent
path'' over the type family \texttt{Gph1\ i\ A\ B\ :\ I\ →\ Set} as
follows:

\begin{Shaded}
\begin{Highlighting}[]
\NormalTok{    lemma0 }\OtherTok{:} \OtherTok{(}\NormalTok{i }\OtherTok{:}\NormalTok{ I}\OtherTok{)} \OtherTok{→}\NormalTok{ Gph1 i A B}
\NormalTok{    lemma0 i }\OtherTok{=}\NormalTok{ α }\OtherTok{(}\NormalTok{Gph1 i A B}\OtherTok{)} \OtherTok{(}\NormalTok{g1pair i a }\OtherTok{(λ} \OtherTok{\_} \OtherTok{→}\NormalTok{ b}\OtherTok{))}
\end{Highlighting}
\end{Shaded}

Then taking the second projection of \texttt{lemma0} evaluated at
\texttt{i1} yields an element of \(B\) evaluated at the first projection
of \texttt{lemma0} evaluated at \texttt{i1}:

\begin{Shaded}
\begin{Highlighting}[]
\NormalTok{    lemma1 }\OtherTok{:}\NormalTok{ B }\OtherTok{(}\NormalTok{g1fst i1 }\OtherTok{(}\NormalTok{lemma0 i1}\OtherTok{))}
\NormalTok{    lemma1 }\OtherTok{=}\NormalTok{ g1snd }\OtherTok{(}\NormalTok{lemma0 i1}\OtherTok{)}
\end{Highlighting}
\end{Shaded}

And we can use \texttt{lemma0} to construct a path from \texttt{α\ A\ a}
to \texttt{g1fst\ i1\ (lemma0\ i1)} as follows:

\begin{Shaded}
\begin{Highlighting}[]
\NormalTok{    lemma2 }\OtherTok{:}\NormalTok{ Path }\OtherTok{(λ} \OtherTok{\_} \OtherTok{→}\NormalTok{ A}\OtherTok{)} \OtherTok{(}\NormalTok{α A a}\OtherTok{)} \OtherTok{(}\NormalTok{g1fst i1 }\OtherTok{(}\NormalTok{lemma0 i1}\OtherTok{))}
\NormalTok{    lemma2 }\OtherTok{=}\NormalTok{ pabs }\OtherTok{(λ}\NormalTok{ i }\OtherTok{→}\NormalTok{ g1fst i }\OtherTok{(}\NormalTok{lemma0 i}\OtherTok{))}
\end{Highlighting}
\end{Shaded}

And then since \texttt{A} is path-discrete, this path becomes an
equality along which we can transport \texttt{lemma1}:

\begin{verbatim}
    substLemma : B (α A a)
    substLemma = transp⁻¹ B (mkInv idToPath pdA lemma2) lemma1
\end{verbatim}

From this substitution lemma, it straightforwardly follows that any
element of \texttt{PolyId} must be extensionally equivalent to the
polymorphic identity function when evaluated at a path-discrete type:

\begin{Shaded}
\begin{Highlighting}[]
\NormalTok{polyId }\OtherTok{:} \OtherTok{∀} \OtherTok{\{}\NormalTok{ℓ}\OtherTok{\}} \OtherTok{(}\NormalTok{A }\OtherTok{:} \DataTypeTok{Set}\NormalTok{ ℓ}\OtherTok{)} \OtherTok{(}\NormalTok{pdA }\OtherTok{:}\NormalTok{ isPathDiscrete A}\OtherTok{)} \OtherTok{(}\NormalTok{a }\OtherTok{:}\NormalTok{ A}\OtherTok{)}
         \OtherTok{→} \OtherTok{(}\NormalTok{α }\OtherTok{:}\NormalTok{ PolyId ℓ}\OtherTok{)} \OtherTok{→}\NormalTok{ α A a ≡ a}
\NormalTok{polyId A pdA a α }\OtherTok{=}\NormalTok{ paramId}\OtherTok{.}\NormalTok{substLemma A pdA }\OtherTok{(λ}\NormalTok{ b }\OtherTok{→}\NormalTok{ b ≡ a}\OtherTok{)}\NormalTok{ a refl α}
\end{Highlighting}
\end{Shaded}

Before we congratulate ourselves for proving this theorem, however, we
ought to reflect on the significance of what we have proved. For we have
proved \emph{only} that the restrictions of elements of \texttt{PolyId}
to path-discrete types are equivalent to that of the polymorphic
identity function. The theorem would then after all be trivial if it
turned out that the only path-discrete types were (e.g.) those
containing at most one element (i.e.~the \emph{mere propositions}, in
the terminology of HoTT). To show that this is not the case, we make use
of our assumption of \emph{connectedness} for \(I\), which we have
already seen implies that every discrete type is path-discrete. To give
a concrete (non-trivial) instance of this, I now show that the type of
Booleans is discrete (hence path-discrete) and use this to test the
canonicity conjecture on a simple example:

\begin{Shaded}
\begin{Highlighting}[]
\KeywordTok{module}\NormalTok{ BoolDiscrete }\KeywordTok{where}
\end{Highlighting}
\end{Shaded}

Showing that \texttt{Bool} is discrete is a simple matter of pattern
matching

\begin{Shaded}
\begin{Highlighting}[]
\NormalTok{    boolIsDisc }\OtherTok{:}\NormalTok{ isDiscrete Bool}
\NormalTok{    boolIsDisc false }\OtherTok{=} \OtherTok{(}\NormalTok{con false , refl}\OtherTok{)}\NormalTok{ , }\OtherTok{λ} \OtherTok{\{} \OtherTok{(}\NormalTok{con false , refl}\OtherTok{)} \OtherTok{→}\NormalTok{ refl}\OtherTok{\}}
\NormalTok{    boolIsDisc true  }\OtherTok{=} \OtherTok{(}\NormalTok{con true  , refl}\OtherTok{)}\NormalTok{ , }\OtherTok{λ} \OtherTok{\{} \OtherTok{(}\NormalTok{con true , refl}\OtherTok{)} \OtherTok{→}\NormalTok{ refl}\OtherTok{\}}
\end{Highlighting}
\end{Shaded}

It follows that \texttt{Bool} is also path-discrete and so the above
parametricity theorem may be applied to it.

\begin{Shaded}
\begin{Highlighting}[]
\NormalTok{    boolIsPDisc }\OtherTok{:}\NormalTok{ isPathDiscrete Bool}
\NormalTok{    boolIsPDisc }\OtherTok{=}\NormalTok{ isDisc→isPDisc boolIsDisc}

\NormalTok{    polyIdBool }\OtherTok{:} \OtherTok{(}\NormalTok{α }\OtherTok{:}\NormalTok{ PolyId lzero}\OtherTok{)} \OtherTok{→} \OtherTok{(}\NormalTok{b }\OtherTok{:}\NormalTok{ Bool}\OtherTok{)} \OtherTok{→}\NormalTok{ α Bool b ≡ b}
\NormalTok{    polyIdBool α b }\OtherTok{=}\NormalTok{ polyId Bool boolIsPDisc b α}
\end{Highlighting}
\end{Shaded}

We can use this to check that, in at least one specific case, the proof
of \texttt{polyId} yields a canonical form (namely \texttt{refl}) when
it is applied to canonical forms:

\begin{Shaded}
\begin{Highlighting}[]
\NormalTok{    shouldBeRefl1 }\OtherTok{:}\NormalTok{ true ≡ true}
\NormalTok{    shouldBeRefl1 }\OtherTok{=}\NormalTok{ polyIdBool }\OtherTok{(λ}\NormalTok{ X }\OtherTok{→} \OtherTok{λ}\NormalTok{ x }\OtherTok{→}\NormalTok{ x}\OtherTok{)}\NormalTok{ true}
\end{Highlighting}
\end{Shaded}

Running Agda's normalization procedure on this term shows that it does
indeed evaluate to \texttt{refl}.

\subsection{Parametricity \& (Higher) Inductive
Types}\label{parametricity-higher-inductive-types}

The foregoing proof of parametricity for the type of the polymorphic
identity function remains ultimately a toy example. To demonstrate the
true power of this approach to parametricity, I turn now to some more
intricate examples of its use, in proving universal properties for
simplified presentations of inductive types and higher inductive types.

In general, it is easy to write down what should be the \emph{recursion}
principle for an inductive type generated by some set of constructors,
but harder (though feasible) to write down the corresponding
\emph{induction} principle. When one begins to consider more complex
generalizations of inductive types, such as higher inductive types,
inductive-inductive types, etc, these difficulties begin to multiply.
What would be ideal would be a way of deriving the induction principle
for an inductive type from its \emph{recursor,} hence requiring only the
latter to be specified as part of the primitive data of the inductive
type. However, in most systems of ordinary dependent type theory this is
generally not possible \cite{Geuvers}. In HoTT, there is one way around
this issue, due to Awodey, Frey \& Speight \cite{Awodey2018}, whereby additional
\emph{naturality} constraints are imposed upon the would-be inductive
type, that serve to make the corresponding induction principle derivable
from the recursor. However, when one goes on to apply this technique to
\emph{higher inductive types,} which may specify constructors not only
for \emph{elements} of a type, but also for instances of its (higher)
identity types, the complexity of these naturality conditions renders
them impractical to work with. The ballooning complexity of these
conditions is an instance of the infamous \emph{coherence problem} in
HoTT, whereby the complexity of coherence conditions for
higher-categorical structures seemingly escapes any simple inductive
definition.

As an alternative, let us consider ways of using \emph{parametricity} to
derive induction principles for inductive and higher inductive types,
starting with the prototypical inductive type, the natural numbers
\(\mathbb{N}\).

First, we define the type of the recursor for \(\mathbb{N}\):

\begin{Shaded}
\begin{Highlighting}[]
\NormalTok{Recℕ }\OtherTok{:}\NormalTok{ Setω}
\NormalTok{Recℕ }\OtherTok{=} \OtherTok{∀} \OtherTok{\{}\NormalTok{ℓ}\OtherTok{\}} \OtherTok{(}\NormalTok{A }\OtherTok{:} \DataTypeTok{Set}\NormalTok{ ℓ}\OtherTok{)} \OtherTok{→}\NormalTok{ A }\OtherTok{→} \OtherTok{(}\NormalTok{A }\OtherTok{→}\NormalTok{ A}\OtherTok{)} \OtherTok{→}\NormalTok{ A}
\end{Highlighting}
\end{Shaded}

We may then postulate the usual constructors and identities for
\(\mathbb{N}\)\footnote{Note that among the identities postulated as
  rewrite rules for \(\mathbb{N}\) is its \(\eta\)-law,
  i.e.~\(\mathsf{rec\mathbb{N}} ~ n ~ \mathbb{N} ~ \mathsf{zero} ~ \mathsf{succ} = n\)
  for all \(n : \mathbb{N}\). This will be important for deriving the
  induction principle for \(\mathbb{N}\).}

\begin{Shaded}
\begin{Highlighting}[]
\KeywordTok{postulate}
\NormalTok{    ℕ }\OtherTok{:} \DataTypeTok{Set₀}
\NormalTok{    zero }\OtherTok{:}\NormalTok{ ℕ}
\NormalTok{    succ }\OtherTok{:}\NormalTok{ ℕ }\OtherTok{→}\NormalTok{ ℕ}
\NormalTok{    recℕ }\OtherTok{:}\NormalTok{ ℕ }\OtherTok{→}\NormalTok{ Recℕ}
\NormalTok{    zeroβ }\OtherTok{:} \OtherTok{∀} \OtherTok{\{}\NormalTok{ℓ}\OtherTok{\}} \OtherTok{(}\NormalTok{A }\OtherTok{:} \DataTypeTok{Set}\NormalTok{ ℓ}\OtherTok{)} \OtherTok{(}\NormalTok{a }\OtherTok{:}\NormalTok{ A}\OtherTok{)} \OtherTok{(}\NormalTok{f }\OtherTok{:}\NormalTok{ A }\OtherTok{→}\NormalTok{ A}\OtherTok{)} \OtherTok{→}\NormalTok{ recℕ zero A a f ≡ a}
    \PreprocessorTok{\{{-}\# REWRITE zeroβ \#{-}\}}
\NormalTok{    succβ }\OtherTok{:} \OtherTok{∀} \OtherTok{\{}\NormalTok{ℓ}\OtherTok{\}} \OtherTok{(}\NormalTok{n }\OtherTok{:}\NormalTok{ ℕ}\OtherTok{)} \OtherTok{(}\NormalTok{A }\OtherTok{:} \DataTypeTok{Set}\NormalTok{ ℓ}\OtherTok{)} \OtherTok{(}\NormalTok{a }\OtherTok{:}\NormalTok{ A}\OtherTok{)} \OtherTok{(}\NormalTok{f }\OtherTok{:}\NormalTok{ A }\OtherTok{→}\NormalTok{ A}\OtherTok{)}
            \OtherTok{→}\NormalTok{ recℕ }\OtherTok{(}\NormalTok{succ n}\OtherTok{)}\NormalTok{ A a f ≡ f }\OtherTok{(}\NormalTok{recℕ n A a f}\OtherTok{)}
    \PreprocessorTok{\{{-}\# REWRITE succβ \#{-}\}}
\NormalTok{    ℕη }\OtherTok{:} \OtherTok{(}\NormalTok{n }\OtherTok{:}\NormalTok{ ℕ}\OtherTok{)} \OtherTok{→}\NormalTok{ recℕ n ℕ zero succ ≡ n}
    \PreprocessorTok{\{{-}\# REWRITE ℕη \#{-}\}}
\end{Highlighting}
\end{Shaded}

From here, we may proceed essentially as in the proof of parametricity
for \texttt{PolyId}, by proving an analogous \emph{substitution lemma}
for \texttt{Recℕ}, following essentially the same steps:

\begin{Shaded}
\begin{Highlighting}[]
\KeywordTok{module}\NormalTok{ paramℕ }\OtherTok{\{}\NormalTok{ℓ}\OtherTok{\}} \OtherTok{(}\NormalTok{α }\OtherTok{:}\NormalTok{ Recℕ}\OtherTok{)} \OtherTok{(}\NormalTok{A }\OtherTok{:} \DataTypeTok{Set}\NormalTok{ ℓ}\OtherTok{)} \OtherTok{(}\NormalTok{pdA }\OtherTok{:}\NormalTok{ isPathDiscrete A}\OtherTok{)} 
                  \OtherTok{(}\NormalTok{B }\OtherTok{:}\NormalTok{ A }\OtherTok{→} \DataTypeTok{Set}\NormalTok{ ℓ}\OtherTok{)} \OtherTok{(}\NormalTok{a }\OtherTok{:}\NormalTok{ A}\OtherTok{)} \OtherTok{(}\NormalTok{b }\OtherTok{:}\NormalTok{ B a}\OtherTok{)}
                  \OtherTok{(}\NormalTok{f }\OtherTok{:}\NormalTok{ A }\OtherTok{→}\NormalTok{ A}\OtherTok{)} \OtherTok{(}\NormalTok{ff }\OtherTok{:} \OtherTok{(}\NormalTok{x }\OtherTok{:}\NormalTok{ A}\OtherTok{)} \OtherTok{→}\NormalTok{ B x }\OtherTok{→}\NormalTok{ B }\OtherTok{(}\NormalTok{f x}\OtherTok{))} \KeywordTok{where}

\NormalTok{    lemma0 }\OtherTok{:} \OtherTok{(}\NormalTok{i }\OtherTok{:}\NormalTok{ I}\OtherTok{)} \OtherTok{→}\NormalTok{ Gph1 i A B}
\NormalTok{    lemma0 i }\OtherTok{=}\NormalTok{ α }\OtherTok{(}\NormalTok{Gph1 i A B}\OtherTok{)}
                 \OtherTok{(}\NormalTok{g1pair i a }\OtherTok{(λ} \OtherTok{\_} \OtherTok{→}\NormalTok{ b}\OtherTok{))}
                 \OtherTok{(λ}\NormalTok{ g }\OtherTok{→} \KeywordTok{let}\NormalTok{ g\textquotesingle{} j q }\OtherTok{=}\NormalTok{ transp }\OtherTok{(λ}\NormalTok{ k }\OtherTok{→}\NormalTok{ Gph1 k A B}\OtherTok{)}\NormalTok{ q g }\KeywordTok{in}
\NormalTok{                        g1pair i }\OtherTok{(}\NormalTok{f }\OtherTok{(}\NormalTok{g1fst i g}\OtherTok{))}
                               \OtherTok{(λ}\NormalTok{ p }\OtherTok{→}\NormalTok{ J⁻¹ }\OtherTok{(λ}\NormalTok{ j q }\OtherTok{→}\NormalTok{ B }\OtherTok{(}\NormalTok{f }\OtherTok{(}\NormalTok{g1fst j }\OtherTok{(}\NormalTok{g\textquotesingle{} j q}\OtherTok{))))}\NormalTok{ p}
                                          \OtherTok{(}\NormalTok{ff }\OtherTok{(}\NormalTok{g1fst i1 }\OtherTok{(}\NormalTok{g\textquotesingle{} i1 p}\OtherTok{))} 
                                              \OtherTok{(}\NormalTok{g1snd }\OtherTok{(}\NormalTok{g\textquotesingle{} i1 p}\OtherTok{)))))}

\NormalTok{    lemma1 }\OtherTok{:}\NormalTok{ B }\OtherTok{(}\NormalTok{g1fst i1 }\OtherTok{(}\NormalTok{lemma0 i1}\OtherTok{))}
\NormalTok{    lemma1 }\OtherTok{=}\NormalTok{ g1snd }\OtherTok{(}\NormalTok{lemma0 i1}\OtherTok{)}

\NormalTok{    lemma2 }\OtherTok{:}\NormalTok{ Path }\OtherTok{(λ} \OtherTok{\_} \OtherTok{→}\NormalTok{ A}\OtherTok{)} \OtherTok{(}\NormalTok{α A a f}\OtherTok{)} \OtherTok{(}\NormalTok{g1fst i1 }\OtherTok{(}\NormalTok{lemma0 i1}\OtherTok{))}
\NormalTok{    lemma2 }\OtherTok{=}\NormalTok{ pabs }\OtherTok{(λ}\NormalTok{ i }\OtherTok{→}\NormalTok{ g1fst i }\OtherTok{(}\NormalTok{lemma0 i}\OtherTok{))}

\NormalTok{    substLemma }\OtherTok{:}\NormalTok{ B }\OtherTok{(}\NormalTok{α A a f}\OtherTok{)}
\NormalTok{    substLemma }\OtherTok{=}\NormalTok{ transp⁻¹ B }\OtherTok{(}\NormalTok{mkInv idToPath pdA lemma2}\OtherTok{)}\NormalTok{ lemma1}
\end{Highlighting}
\end{Shaded}

In order to apply this lemma to \(\mathbb{N}\) itself, we must further
postulate that \(\mathbb{N}\) is path-discrete (in fact, one could show
by induction that \(ℕ\) is \emph{discrete}, hence path-discrete by the
assumption of connectedness for \(I\); however, since we have not yet
proven induction for \(\mathbb{N}\), we must instead take this result as
an additional axiom, from which induction on \(\mathbb{N}\) will
follow).

\begin{Shaded}
\begin{Highlighting}[]
\KeywordTok{postulate}
\NormalTok{    pdℕ1 }\OtherTok{:} \OtherTok{∀} \OtherTok{\{}\NormalTok{m n }\OtherTok{:}\NormalTok{ ℕ}\OtherTok{\}} \OtherTok{(}\NormalTok{e }\OtherTok{:}\NormalTok{ Path }\OtherTok{(λ} \OtherTok{\_} \OtherTok{→}\NormalTok{ ℕ}\OtherTok{)}\NormalTok{ m n}\OtherTok{)} 
           \OtherTok{→}\NormalTok{ Σ }\OtherTok{(}\NormalTok{m ≡ n}\OtherTok{)} \OtherTok{(λ}\NormalTok{ p }\OtherTok{→}\NormalTok{ idToPath p ≡ e}\OtherTok{)}
\NormalTok{    pdℕ2 }\OtherTok{:} \OtherTok{∀} \OtherTok{\{}\NormalTok{m n }\OtherTok{:}\NormalTok{ ℕ}\OtherTok{\}} \OtherTok{(}\NormalTok{e }\OtherTok{:}\NormalTok{ Path }\OtherTok{(λ} \OtherTok{\_} \OtherTok{→}\NormalTok{ ℕ}\OtherTok{)}\NormalTok{ m n}\OtherTok{)}
           \OtherTok{→} \OtherTok{(}\NormalTok{q }\OtherTok{:}\NormalTok{ m ≡ n}\OtherTok{)} \OtherTok{(}\NormalTok{r }\OtherTok{:}\NormalTok{ idToPath q ≡ e}\OtherTok{)}
           \OtherTok{→}\NormalTok{ pdℕ1 e ≡ }\OtherTok{(}\NormalTok{q , r}\OtherTok{)}

\NormalTok{pdℕ }\OtherTok{:}\NormalTok{ isPathDiscrete ℕ}
\NormalTok{pdℕ e }\OtherTok{=} \OtherTok{(}\NormalTok{pdℕ1 e , }\OtherTok{λ} \OtherTok{(}\NormalTok{q , r}\OtherTok{)} \OtherTok{→}\NormalTok{ pdℕ2 e q r}\OtherTok{)}

\NormalTok{rwPDℕ1 }\OtherTok{:} \OtherTok{(}\NormalTok{n }\OtherTok{:}\NormalTok{ ℕ}\OtherTok{)} \OtherTok{→}\NormalTok{ pdℕ1 }\OtherTok{(}\NormalTok{pabs }\OtherTok{(λ} \OtherTok{\_} \OtherTok{→}\NormalTok{ n}\OtherTok{))}\NormalTok{ ≡ }\OtherTok{(}\NormalTok{refl , refl}\OtherTok{)}
\NormalTok{rwPDℕ1 n }\OtherTok{=}\NormalTok{ pdℕ2 }\OtherTok{(}\NormalTok{pabs }\OtherTok{(λ} \OtherTok{\_} \OtherTok{→}\NormalTok{ n}\OtherTok{))}\NormalTok{ refl refl}
\PreprocessorTok{\{{-}\# REWRITE rwPDℕ1 \#{-}\}}

\KeywordTok{postulate}
\NormalTok{    rwPDℕ2 }\OtherTok{:} \OtherTok{(}\NormalTok{n }\OtherTok{:}\NormalTok{ ℕ}\OtherTok{)} \OtherTok{→}\NormalTok{ pdℕ2 }\OtherTok{(}\NormalTok{pabs }\OtherTok{(λ} \OtherTok{\_} \OtherTok{→}\NormalTok{ n}\OtherTok{))}\NormalTok{ refl refl ≡ refl}
    \PreprocessorTok{\{{-}\# REWRITE rwPDℕ2 \#{-}\}}
\end{Highlighting}
\end{Shaded}

Induction for \(\mathbb{N}\) then follows as a straightforward
consequence of the substitution lemma:\footnote{Note that our use of the
  \(\eta\)-law for \(\mathbb{N}\) as a rewrite rule is critical to the
  above proof, since otherwise in the last step, we would obtain not a
  proof of \texttt{P\ n}, but rather
  \texttt{P\ (recℕ\ n\ ℕ\ succ\ zero)}.}

\begin{Shaded}
\begin{Highlighting}[]
\NormalTok{indℕ }\OtherTok{:} \OtherTok{(}\NormalTok{P }\OtherTok{:}\NormalTok{ ℕ }\OtherTok{→} \DataTypeTok{Set}\OtherTok{)} \OtherTok{→}\NormalTok{ P zero }\OtherTok{→} \OtherTok{((}\NormalTok{n }\OtherTok{:}\NormalTok{ ℕ}\OtherTok{)} \OtherTok{→}\NormalTok{ P n }\OtherTok{→}\NormalTok{ P }\OtherTok{(}\NormalTok{succ n}\OtherTok{))} \OtherTok{→} \OtherTok{(}\NormalTok{n }\OtherTok{:}\NormalTok{ ℕ}\OtherTok{)} \OtherTok{→}\NormalTok{ P n}
\NormalTok{indℕ P pz ps n }\OtherTok{=}\NormalTok{ paramℕ}\OtherTok{.}\NormalTok{substLemma }\OtherTok{(}\NormalTok{recℕ n}\OtherTok{)}\NormalTok{ ℕ pdℕ P zero pz succ ps}
\end{Highlighting}
\end{Shaded}

As in the case of the parametricity theorem for \texttt{PolyId}, we may
test that the derived induction principle for \(\mathbb{N}\) evaluates
canonical forms to canonical forms. The following example tests this for
the usual inductive proof that \texttt{zero} is an identity element for
addition on the right:

\begin{Shaded}
\begin{Highlighting}[]
\KeywordTok{module}\NormalTok{ ℕexample }\KeywordTok{where}
    \OtherTok{\_}\NormalTok{plus}\OtherTok{\_} \OtherTok{:}\NormalTok{ ℕ }\OtherTok{→}\NormalTok{ ℕ }\OtherTok{→}\NormalTok{ ℕ}
\NormalTok{    m plus n }\OtherTok{=}\NormalTok{ recℕ m ℕ n succ}

\NormalTok{    zeroIdR }\OtherTok{:} \OtherTok{(}\NormalTok{n }\OtherTok{:}\NormalTok{ ℕ}\OtherTok{)} \OtherTok{→}\NormalTok{ n plus zero ≡ n}
\NormalTok{    zeroIdR n }\OtherTok{=}\NormalTok{ indℕ }\OtherTok{(λ}\NormalTok{ m }\OtherTok{→}\NormalTok{ m plus zero ≡ m}\OtherTok{)}\NormalTok{ refl }\OtherTok{(λ}\NormalTok{ m p }\OtherTok{→}\NormalTok{ ap succ p}\OtherTok{)}\NormalTok{ n}
    
\NormalTok{    shouldBeRefl2 }\OtherTok{:}\NormalTok{ succ }\OtherTok{(}\NormalTok{succ zero}\OtherTok{)}\NormalTok{ ≡ succ }\OtherTok{(}\NormalTok{succ zero}\OtherTok{)}
\NormalTok{    shouldBeRefl2 }\OtherTok{=}\NormalTok{ zeroIdR }\OtherTok{(}\NormalTok{succ }\OtherTok{(}\NormalTok{succ zero}\OtherTok{))}
\end{Highlighting}
\end{Shaded}

Running Agda's normalization procedure on \texttt{shouldBeRefl2} again
confirms that it evaluates to \texttt{refl}.

Moving on, then, from inductive types to \emph{higher} inductive types,
we may now consider deriving the induction principle for the
\emph{circle} \(S^1\), defined to be the type freely generated by a
single basepoint \(\mathsf{base} : S^1\), with a nontrivial
identification \(\mathsf{loop} : \mathsf{base} = \mathsf{base}\). The
recursor for \(S^1\) thus has the following type:

\begin{Shaded}
\begin{Highlighting}[]
\NormalTok{RecS¹ }\OtherTok{:}\NormalTok{ Setω}
\NormalTok{RecS¹ }\OtherTok{=} \OtherTok{∀} \OtherTok{\{}\NormalTok{ℓ}\OtherTok{\}} \OtherTok{(}\NormalTok{A }\OtherTok{:} \DataTypeTok{Set}\NormalTok{ ℓ}\OtherTok{)} \OtherTok{→} \OtherTok{(}\NormalTok{a }\OtherTok{:}\NormalTok{ A}\OtherTok{)} \OtherTok{→}\NormalTok{ a ≡ a }\OtherTok{→}\NormalTok{ A}
\end{Highlighting}
\end{Shaded}

Then, just as before, we may postulate the corresponding constructors
and \(\beta\eta\)-laws for \(S^1\):

\begin{Shaded}
\begin{Highlighting}[]
\KeywordTok{postulate}
\NormalTok{    S¹ }\OtherTok{:} \DataTypeTok{Set₀}
\NormalTok{    base }\OtherTok{:}\NormalTok{ S¹}
\NormalTok{    loop }\OtherTok{:}\NormalTok{ base ≡ base}
\NormalTok{    recS¹ }\OtherTok{:}\NormalTok{ S¹ }\OtherTok{→}\NormalTok{ RecS¹}
\NormalTok{    baseβ }\OtherTok{:} \OtherTok{∀} \OtherTok{\{}\NormalTok{ℓ}\OtherTok{\}} \OtherTok{(}\NormalTok{A }\OtherTok{:} \DataTypeTok{Set}\NormalTok{ ℓ}\OtherTok{)} \OtherTok{(}\NormalTok{a }\OtherTok{:}\NormalTok{ A}\OtherTok{)} \OtherTok{(}\NormalTok{l }\OtherTok{:}\NormalTok{ a ≡ a}\OtherTok{)} \OtherTok{→}\NormalTok{ recS¹ base A a l ≡ a}
    \PreprocessorTok{\{{-}\# REWRITE baseβ \#{-}\}}
\NormalTok{    loopβ }\OtherTok{:} \OtherTok{∀} \OtherTok{\{}\NormalTok{ℓ}\OtherTok{\}} \OtherTok{(}\NormalTok{A }\OtherTok{:} \DataTypeTok{Set}\NormalTok{ ℓ}\OtherTok{)} \OtherTok{(}\NormalTok{a }\OtherTok{:}\NormalTok{ A}\OtherTok{)} \OtherTok{(}\NormalTok{l }\OtherTok{:}\NormalTok{ a ≡ a}\OtherTok{)}
              \OtherTok{→}\NormalTok{ ap }\OtherTok{(λ}\NormalTok{ s }\OtherTok{→}\NormalTok{ recS¹ s A a l}\OtherTok{)}\NormalTok{ loop ≡ l}
    \PreprocessorTok{\{{-}\# REWRITE loopβ \#{-}\}}
\NormalTok{    S¹η }\OtherTok{:} \OtherTok{(}\NormalTok{s }\OtherTok{:}\NormalTok{ S¹}\OtherTok{)} \OtherTok{→}\NormalTok{ recS¹ s S¹ base loop ≡ s}
    \PreprocessorTok{\{{-}\# REWRITE S¹η \#{-}\}}
\end{Highlighting}
\end{Shaded}

The proof of induction for \(S^1\) is then in essentials the same as the
one given above for \(\mathbb{N}\). We begin by proving a
\emph{substitution lemma} for \texttt{RecS¹}, following exactly the same
steps as in the proof of the corresponding theorem for \texttt{Recℕ}:

\begin{Shaded}
\begin{Highlighting}[]
\KeywordTok{module}\NormalTok{ paramS¹ }\OtherTok{\{}\NormalTok{ℓ}\OtherTok{\}} \OtherTok{(}\NormalTok{A }\OtherTok{:} \DataTypeTok{Set}\NormalTok{ ℓ}\OtherTok{)} \OtherTok{(}\NormalTok{pdA }\OtherTok{:}\NormalTok{ isPathDiscrete A}\OtherTok{)} \OtherTok{(}\NormalTok{B }\OtherTok{:}\NormalTok{ A }\OtherTok{→} \DataTypeTok{Set}\NormalTok{ ℓ}\OtherTok{)} 
                   \OtherTok{(}\NormalTok{a }\OtherTok{:}\NormalTok{ A}\OtherTok{)} \OtherTok{(}\NormalTok{b }\OtherTok{:}\NormalTok{ B a}\OtherTok{)} \OtherTok{(}\NormalTok{l }\OtherTok{:}\NormalTok{ a ≡ a}\OtherTok{)}
                   \OtherTok{(}\NormalTok{lB }\OtherTok{:}\NormalTok{ b ≡ transp⁻¹ B l b}\OtherTok{)} \OtherTok{(}\NormalTok{α }\OtherTok{:}\NormalTok{ RecS¹}\OtherTok{)} \KeywordTok{where}

\NormalTok{    lemma0 }\OtherTok{:} \OtherTok{(}\NormalTok{i }\OtherTok{:}\NormalTok{ I}\OtherTok{)} \OtherTok{→}\NormalTok{ Gph1 i A B}
\NormalTok{    lemma0 i }\OtherTok{=}\NormalTok{ α }\OtherTok{(}\NormalTok{Gph1 i A B}\OtherTok{)} \OtherTok{(}\NormalTok{g1pair i a }\OtherTok{(λ} \OtherTok{\_} \OtherTok{→}\NormalTok{ b}\OtherTok{))} \OtherTok{(}\NormalTok{apg1pair l lB i}\OtherTok{)}

\NormalTok{    lemma1 }\OtherTok{:}\NormalTok{ B }\OtherTok{(}\NormalTok{g1fst i1 }\OtherTok{(}\NormalTok{lemma0 i1}\OtherTok{))}
\NormalTok{    lemma1 }\OtherTok{=}\NormalTok{ g1snd }\OtherTok{(}\NormalTok{lemma0 i1}\OtherTok{)}

\NormalTok{    lemma2 }\OtherTok{:}\NormalTok{ Path }\OtherTok{(λ} \OtherTok{\_} \OtherTok{→}\NormalTok{ A}\OtherTok{)} \OtherTok{(}\NormalTok{α A a l}\OtherTok{)} \OtherTok{(}\NormalTok{g1fst i1 }\OtherTok{(}\NormalTok{lemma0 i1}\OtherTok{))}
\NormalTok{    lemma2 }\OtherTok{=}\NormalTok{ pabs }\OtherTok{(λ}\NormalTok{ i }\OtherTok{→}\NormalTok{ g1fst i }\OtherTok{(}\NormalTok{lemma0 i}\OtherTok{))}

\NormalTok{    substLemma }\OtherTok{:}\NormalTok{ B }\OtherTok{(}\NormalTok{α A a l}\OtherTok{)}
\NormalTok{    substLemma }\OtherTok{=}\NormalTok{ transp⁻¹ B }\OtherTok{(}\NormalTok{mkInv idToPath pdA lemma2}\OtherTok{)}\NormalTok{ lemma1}
\end{Highlighting}
\end{Shaded}

We then postulate that \(S^1\) is path-discrete, as before, in order to
apply this lemma to \(S^1\) itself:

\begin{Shaded}
\begin{Highlighting}[]
\KeywordTok{postulate}
\NormalTok{    pdS¹1 }\OtherTok{:} \OtherTok{∀} \OtherTok{\{}\NormalTok{s t }\OtherTok{:}\NormalTok{ S¹}\OtherTok{\}} \OtherTok{(}\NormalTok{e }\OtherTok{:}\NormalTok{ Path }\OtherTok{(λ} \OtherTok{\_} \OtherTok{→}\NormalTok{ S¹}\OtherTok{)}\NormalTok{ s t}\OtherTok{)}
            \OtherTok{→}\NormalTok{ Σ }\OtherTok{(}\NormalTok{s ≡ t}\OtherTok{)} \OtherTok{(λ}\NormalTok{ p }\OtherTok{→}\NormalTok{ idToPath p ≡ e}\OtherTok{)}
\NormalTok{    pdS¹2 }\OtherTok{:} \OtherTok{∀} \OtherTok{\{}\NormalTok{s t }\OtherTok{:}\NormalTok{ S¹}\OtherTok{\}} \OtherTok{(}\NormalTok{e }\OtherTok{:}\NormalTok{ Path }\OtherTok{(λ} \OtherTok{\_} \OtherTok{→}\NormalTok{ S¹}\OtherTok{)}\NormalTok{ s t}\OtherTok{)}
            \OtherTok{→} \OtherTok{(}\NormalTok{q }\OtherTok{:}\NormalTok{ s ≡ t}\OtherTok{)} \OtherTok{(}\NormalTok{r }\OtherTok{:}\NormalTok{ idToPath q ≡ e}\OtherTok{)}
            \OtherTok{→}\NormalTok{ pdS¹1 e ≡ }\OtherTok{(}\NormalTok{q , r}\OtherTok{)}

\NormalTok{pdS¹ }\OtherTok{:}\NormalTok{ isPathDiscrete S¹}
\NormalTok{pdS¹ e }\OtherTok{=} \OtherTok{(}\NormalTok{pdS¹1 e , }\OtherTok{λ} \OtherTok{(}\NormalTok{q , r}\OtherTok{)} \OtherTok{→}\NormalTok{ pdS¹2 e q r}\OtherTok{)}

\NormalTok{rwPDS¹1 }\OtherTok{:} \OtherTok{(}\NormalTok{s }\OtherTok{:}\NormalTok{ S¹}\OtherTok{)} \OtherTok{→}\NormalTok{ pdS¹1 }\OtherTok{(}\NormalTok{pabs }\OtherTok{(λ} \OtherTok{\_} \OtherTok{→}\NormalTok{ s}\OtherTok{))}\NormalTok{ ≡ }\OtherTok{(}\NormalTok{refl , refl}\OtherTok{)}
\NormalTok{rwPDS¹1 s }\OtherTok{=}\NormalTok{ pdS¹2 }\OtherTok{(}\NormalTok{pabs }\OtherTok{(λ} \OtherTok{\_} \OtherTok{→}\NormalTok{ s}\OtherTok{))}\NormalTok{ refl refl}
\PreprocessorTok{\{{-}\# REWRITE rwPDS¹1 \#{-}\}}

\KeywordTok{postulate}
\NormalTok{    rwPDS¹2 }\OtherTok{:} \OtherTok{(}\NormalTok{s }\OtherTok{:}\NormalTok{ S¹}\OtherTok{)} \OtherTok{→}\NormalTok{ pdS¹2 }\OtherTok{(}\NormalTok{pabs }\OtherTok{(λ} \OtherTok{\_} \OtherTok{→}\NormalTok{ s}\OtherTok{))}\NormalTok{ refl refl ≡ refl}
    \PreprocessorTok{\{{-}\# REWRITE rwPDS¹2 \#{-}\}}
\end{Highlighting}
\end{Shaded}

And then the desired induction principle for \(S^1\) follows
straightforwardly:

\begin{Shaded}
\begin{Highlighting}[]
\NormalTok{indS¹ }\OtherTok{:} \OtherTok{(}\NormalTok{P }\OtherTok{:}\NormalTok{ S¹ }\OtherTok{→} \DataTypeTok{Set}\OtherTok{)} \OtherTok{(}\NormalTok{pb }\OtherTok{:}\NormalTok{ P base}\OtherTok{)} \OtherTok{→}\NormalTok{ pb ≡ transp⁻¹ P loop pb }\OtherTok{→} \OtherTok{(}\NormalTok{s }\OtherTok{:}\NormalTok{ S¹}\OtherTok{)} \OtherTok{→}\NormalTok{ P s}
\NormalTok{indS¹ P pb pl s }\OtherTok{=}\NormalTok{ paramS¹}\OtherTok{.}\NormalTok{substLemma S¹ pdS¹ P base pb loop pl }\OtherTok{(}\NormalTok{recS¹ s}\OtherTok{)}
\end{Highlighting}
\end{Shaded}

Although it is not in general possible to verify that this same
construction is capable of deriving induction principles for \emph{all}
higher inductive types -- essentially because there is as yet no
well-established definition of what higher inductive types are \emph{in
general} -- there appears to be no difficulty in extending this method
of proof to all known classes of higher inductive types. Moreover, that
the proof of induction for \(S^1\) is essentially no more complex than
that for \(\mathbb{N}\) suggests that this method is capable of taming
the complexity of coherences for such higher inductive types, and in
this sense provides a solution to this instance of the coherence
problem.

\section{Toward a synthetic theory of
parametricity}\label{toward-a-synthetic-theory-of-parametricity}

The theory so-far developed essentially gives a synthetic framework for
working in the internal language of a (weakly) \emph{sufficiently
cohesive \(\infty\)-topos}. This framework in turn proves capable of
deriving significant parametricity results internally, with immediate
applications in e.g.~resolving coherence problems having to do with
higher inductive types. It remains to be seen what further applications
can be developed for this theory and its particular approach to
parametricity. From this perspective, it is profitable to survey what
other approaches there are to internalizing parametricity theorems in
dependent type theory, and how they might be related to the one given in
this paper.

\subsection{Cohesion \& Gluing}\label{cohesion-gluing}

In recent years, there has been some related work toward a synthetic
theory of parametricity in terms of the topos-theoretic construction of
\emph{Artin Gluing}. This approach, outlined initially by Sterling in
his thesis \cite{SterlingThesis} and subsequently spearheaded by Sterling and his
collaborators as part of the more general programme of \emph{Synthetic
Tait Computability} (STC), works in the internal language of a topos
equipped with two (mere) propositions \(L\) and \(R\), that are
\emph{mutually exclusive} in that \(L \wedge R \to \bot\). The central
idea of this approach is to synthetically reconstruct the usual account
of parametricity in terms of logical relations by careful use of the
\emph{open} and \emph{closed} modalities induced by these propositions,
which play a similar role in STC to the \emph{graph types} introduced
above. Using this approach, one can e.g.~prove representation
independence theorems for module signatures as in recent work by Sterling \& Harper \cite{Sterling2021}.

Clearly, there is some affinity between the approach to parametricity in
terms of gluing and that in terms of cohesion presented above, not least
of which having to do with the fact that they both offer
characteristically \emph{modal} perspectives on parametricity. Yet there
is a further sense in which these two approaches are related, which is
that every sufficiently cohesive topos contains a model of Sterling's
setup of STC for parametricity, as follows: given a sufficiently
cohesive topos \(\mathcal{E}\) over some base topos \(\mathcal{S}\), for
any strictly bipointed object \(I \in \mathcal{E}\) with distinguished
points \(i_0, i_1 : I\), the \emph{slice topos} \(\mathcal{E}/I\), whose
internal language corresponds to that of \(\mathcal{E}\) extended with
an arbitrary element \(i : I\), is thereby equipped with two mutually
exclusive propositions, namely \(i = i_0\) and \(i = i_1\). Hence all of
the parametricity theorems available in STC can be recovered in the
above-given framework (in particular, instances of closed modalities can
be encoded as higher inductive types, which, as we have just seen, are
easily added to the above framework).

On the other hand, there is no analogue in STC of the axiom of
connectedness for the interval, and its consequent parametricity
theorems, most notably the above-mentioned derivability of induction
principles for higher inductive types. This suggests that in these
latter results, the structure of \emph{cohesion} and its relation to
parametricity plays an essential role. Nonetheless, it remains
interesting to consider how parametricity via cohesion and parametricity
via gluing may yet prove to be related, and one may hope for a fruitful
cross-pollination between these two theories.

\subsection{Cohesion \& Coherence}\label{cohesion-coherence}

As we have seen, there appears to be an intimate link between
\emph{parametricity}, \emph{cohesion}, and \emph{coherence},
demonstrated (e.g.) by the above-given proof of induction for \(S^1\). The
existence of such a link is further supported by other recent
developments in HoTT and related fields. E.g. Cavallo \& Harper \cite{Cavallo2020} have
used their system of internal parametricity for Cubical Type Theory to
derive coherence theorems for the smash product. More recently still,
Kolomatskaia \& Shulman \cite{kolomatskaia2024} have introduced their system of \emph{Displayed}
Type Theory, which utilizes yet another form of internal parametricity to solve the previously-open problem of representing
\emph{semi-simplicial types} in HoTT.

In outlining the above framework for parametricity in terms of cohesion,
I hope to have taken a first step toward the unification of these
various systems that use parametricity to tackle instances of the
coherence problem. Cavallo \& Harper's system is readily assimilated to
this framework, as are other related systems that take a \emph{cubical}
approach to internal parametricity, such as that of Nuyts, Devriese \&
Vezzossi \cite{NDV}, since these all take their semantics in various topoi of
bicubical sets, which all are sufficiently cohesive over corresponding
topoi of cubical sets. On the other hand, Kolomatskaia \& Shulman's
system cannot be assimilated in quite the same way, since their system
takes its semantics in the \(\infty\)-topos of augmented semisimplicial
spaces, which is \emph{not} cohesive over spaces. It thus remains to be
seen if the above framework can be generalized so as to be inclusive of
this example as well. Alternatively, one might seek to generalize or
modify Kolomatskaia \& Shulman's system so as to be interpretable in an
\(\infty\)-topos that \emph{is} cohesive over spaces, e.g.~the
\(\infty\)-topos of simplicial spaces. If this latter proves feasible,
then this in turn would reveal yet further connections between
parametricity via cohesion and another prominent attempt at a solution
to the coherence problem, namely Riehl \& Shulman's \emph{Simplicial
Type Theory}, which takes its semantics in simplicial spaces \cite{RiehlShulman}.

It appears, in all these cases, that parametricity is \emph{the} tool
for the job of taming the complexity of higher coherences in HoTT and
elsewhere. In this sense, Reynolds was right in thinking that
parametricity captures a fundamental property of abstraction, for, as
any type theorist worth their salt knows, abstraction is ultimately the
best tool we have for managing complexity.

\section*{Acknowledgement}\label{acknowledgements} 

The origins of this paper trace to research I did as an undergraduate at
Merton College, Oxford, in the Summer of 2020, as part of the Merton
College Summer Projects Scheme. Naturally, this work was conducted at a
time of considerable stress for myself and the world at large, and I am
massively indebted to the academic support staff at Merton, who were
immensely helpful in keeping me afloat at that time. I am particularly
grateful to Katy Fifield, Jemma Underdown, and Jane Gover for the
support they provided to me in the course of my undergraduate studies.
More recently, I am grateful to Frank Pfenning and Steve Awodey for
their encouragement in continuing to pursue this line of research.

\bibliographystyle{alpha}
\bibliography{example}

\end{document}